\documentclass[fleqn,usenatbib]{mnras}

\usepackage{newtxtext,newtxmath}
\usepackage[T1]{fontenc}

\DeclareRobustCommand{\VAN}[3]{#2}
\let\VANthebibliography\thebibliography
\def\thebibliography{\DeclareRobustCommand{\VAN}[3]{##3}\VANthebibliography}

\usepackage{graphicx}	% Including figure files
\usepackage{amsmath}	% Advanced maths commands
\usepackage{newtxtext,newtxmath}
\usepackage{comment}
\usepackage{bm}
\usepackage{booktabs}  % for well-spaced horizontal rules
\providecommand{\abs}[1]{\lvert#1\rvert}

\usepackage{caption}
\usepackage[flushleft]{threeparttable} % for "ThreePartTable" environment
\usepackage{xspace}

\usepackage{ulem}

\newcommand{\msun}{M$_{\sun}$\xspace}

\newcommand{\dotom}{$\dot{\omega}$}

\newcommand{\mearth}{M$_{\oplus}$}

\newcommand{\dev}{\mathrm{d}}

\newcommand{\ENTERPRISE}{\texttt{ENTERPRISE}}
\newcommand{\TEMPOTWO}{\texttt{TEMPO2}\xspace}

\newcommand{\RNum}[1]{\uppercase\expandafter{\romannumeral #1\relax}}

\title[Timing of J1906]{Long-term timing of the relativistic binary PSR~J1906+0746}

\author[L. Vleeschower et al.]{\parbox{\textwidth}{
L.~Vleeschower,$^{1,2}$\thanks{E-mail: laila.vleeschowercalas@postgrad.manchester.ac.uk}
B.~W.~Stappers,$^{1}$ 
M.~J.~Keith,$^{1}$
G.~Desvignes,$^{3}$
P.~C.~C.~Freire,$^{3}$
M.~Kramer,$^{3,1}$
J.~van Leeuwen,$^{4}$
L.~Levin,$^{1}$
A.~G.~Lyne,$^{1}$
I.~H.~Stairs,${^5}$
V.~Venkatraman~Krishnan,$^{3}$
Y.~Y.~Wang$^{6}$
}
\\ \\ \\ \\
\parbox{\textwidth}{
$^{1}$Jodrell Bank Centre for Astrophysics, Department of Physics and Astronomy, The University of Manchester, Manchester M13 9PL, UK\\
$^{2}$Center for Gravitation, Cosmology, and Astrophysics, Department of Physics, University of Wisconsin-Milwaukee, P.O. Box 413, Milwaukee, WI 53201, USA\\
$^{3}$Max-Planck-Institut f\"{u}r Radioastronomie, Auf dem H\"{u}gel 69, D-53121 Bonn, Germany\label{mpifr}\\
$^{4}$ASTRON, the Netherlands Institute for Radio Astronomy, Oude Hoogeveensedijk 4, 7991 PD Dwingeloo, The Netherlands \\
$^{5}$Dept. of Physics and Astronomy, University of British Columbia, 6224 Agricultural Road, Vancouver, BC V6T 1Z1 Canada\\
$^{6}$Anton Pannekoek Institute for Astronomy, University of Amsterdam, Science Park 904, 1098 XH Amsterdam, The Netherlands\\
}
}

\date{Accepted XXX. Received YYY; in original form ZZZ}

\pubyear{2026}

\begin{document}
\label{firstpage}
\pagerange{\pageref{firstpage}--\pageref{lastpage}}
\maketitle

\begin{abstract}

We conducted a timing analysis of over 18 years of data on the young ($\tau_{\rm c} = 112$ kyr, $P = 114$\,ms) relativistic binary PSR~J1906+0746, using six radio telescopes: Arecibo, FAST, Green Bank, Lovell, MeerKAT, and Nançay. This pulsar is known to orbit a compact high-mass companion with a period of 3.98\,hrs in a mildly eccentric orbit ($e = 0.085$). By combining all data and maintaining a coherent timing solution over the full span, we obtained a more precise measurement of the advance of periastron, $\dot{\omega} = 7.5841(2)$\,$\deg$\,yr$^{-1}$, the Einstein delay, $\gamma = 4.59(2) \times 10^{-4}$\,s, and the secular change in orbital period, $\dot{P}_{\rm b} = -5.65(2) \times 10^{-13}$\,s\,s$^{-1}$. Assuming the validity of general relativity, we obtain a total mass of 2.6133(1)\,$M_{\odot}$ and component masses of 1.316(5) $M_{\odot}$ for the pulsar and 1.297(5) $M_{\odot}$ for the companion, consistent with a double neutron star system. However, when fitting for the secular change in the projected semi-major axis we obtain $\dot{x} = -1.8(6) \times 10^{-13}$\,s\,s$^{-1}$, the component masses are shifted by $\sim 3.5\sigma$, which is expected from the correlation of $\dot{x}$ and $\gamma$. The $\dot{x}$ has a similar magnitude to that observed in PSR~J1141$-$6545, which is due to spin-orbit coupling; if confirmed, it would indicate that, as in the latter system, the companion of the PSR~J1906+0746 system is a massive fast-rotating white dwarf formed before the pulsar.
Additionally, we report and characterize a large glitch near MJD 56664, with a fractional frequency increase comparable to those observed in the Vela pulsar.

\end{abstract}


%%%%%%%%%%%%%%%%%%%%%%%%%%%%%%%%%%%%%%%%%%%%%%%%%%

%%%%%%%%%%%%%%%%% BODY OF PAPER %%%%%%%%%%%%%%%%%%

%%%%%%%%%%%%%%%%%%%% Introduction %%%%%%%%%%%%%%%%%%

\section{Introduction}

PSR~J1906$+$0746, with a spin period of 144\,ms, was discovered in data taken during the pulsar survey observations with the Arecibo L-band Feed Array (P-ALFA) system \citep{Cordes+2006} on 2004 September 27. The pulsar was later also found in archival data from the Parkes Multibeam Pulsar Survey \citep[PMPS;][]{Lorimer+2006} displaying a high degree of acceleration and indicating that PSR~J1906+0746 is part of a short-period binary system. 

A refinement of the orbital and spin parameters was made possible through follow-up observations made with the Arecibo, Green Bank, Lovell, and Murriyang, CSIRO's Parkes radio telescope, indicating a short orbital period ($P_{\rm b} \sim$3.98\,hrs) and a moderate eccentricity ($e \sim 0.085$). The spin frequency ($\nu$) and the rate of change in the spin frequency ($\dot{\nu}$) suggested that this is a young pulsar with a characteristic age ($\tau_{\rm c}$) of only 112 kyr and a strong magnetic field $B = 1.7 \times 10^{12}$\,G \citep{Lorimer+2006,vanLeeuwen+2006}. This makes it the binary pulsar with the smallest known characteristic age\footnote{\url{https://www.atnf.csiro.au/research/pulsar/psrcat/}} and, at the time, the second-shortest orbital period, surpassed only by the double pulsar PSR~J0737--3039 \citep[see e.g.][]{vanLeeuwen+2015, Kramer+2021}. 

PSR~J1906+0746 has been intensively monitored since its discovery. In \citet{vanLeeuwen+2015}, all the data available at that time from the Arecibo, Green Bank, Lovell, Nançay, and Westerbork telescopes were used, spanning from 2005 May until the end of 2009 and comprising 28,000 times of arrival (ToAs). By removing the timing noise present using \texttt{FITWAVES} from \TEMPOTWO\footnote{\url{https://bitbucket.org/psrsoft/tempo2/src/master/}} \citep[see e.g.,][]{Hobbs+2004} and fitting for arbitrary offsets around every set of ToAs, the authors were able to put constraints on the masses of the system by using the estimates of three post-Keplerian parameters: the advance of periastron \dotom\,\,$= 7.5841(5)$\,$\deg$\,yr$^{-1}$, the time dilation parameter or ``Einstein delay'' $\gamma = 4.70(5) \times 10^{-4}$\,s, and the orbital decay $\dot{P_{\rm b}} = -5.6(3) \times 10^{-13}$. In \citet{vanLeeuwen+2015} the pulsar mass was concluded to be $M_{\rm p} = 1.291(11)$\,\msun and the companion mass was $M_{\rm c} = 1.322(11)$\,\msun. These masses fit the observed sample and the standard evolutionary model of double neutron star (DNS) systems.

However, the pulsar is not recycled, indicating that it is the second-formed compact object in the system. This implies that, in principle, the first-formed compact object in the system could be a massive white dwarf \citep[WD;][]{Tauris+Sennels2000}, as in the case of similar systems like PSR~B2303+46 \citep{vanKerkwijk+Kulkarni1999} and especially PSR~J1141--6545, which also has a similar orbital period \citep{Kaspi+2000} and for which the WD nature of the companion has also been confirmed \citep{Antoniadis+2011}.
Furthermore, the companion mass lies within the mass range of well-measured WD masses in those systems as well as the precisely measured mass of the WD companion of PSR~J2222--0137 \citep{Guo+2021}.

PSR~J1906+0746 is known to display a significant amount of linear polarisation \citep{Lorimer+2006, Desvignes+2008, Desvignes+2019} as seen for other young pulsars \citep[see e.g.,][]{Weltevrede+Johnston2008}. Moreover, no mode changes have been detected on any time scales from single-pulse data to 10, 30, 60, and 120\,s integrations \citep{vanLeeuwen+2015}. The pulse profile of this pulsar has been seen to evolve over timescales of years. The interpulse seen in the 2004 discovery pulse profile was absent in the 1998 Parkes archive profile \citep{Lorimer+2006,vanLeeuwen+2006}. Furthermore, the separation between the main pulse and the interpulse has also been seen to change over time. The profile evolution can be attributed to the general relativistic effect of geodetic precession, in which the pulsar's spin angular momentum vector precesses around the total system's angular momentum \citep{Desvignes+2019}. The profile variations occur as the precessing pulsar beam changes its orientation with respect to our line of sight.

The presence of emission from both the main pulse and the interpulse covering a wide range of pulse longitudes has allowed a precise determination of the viewing geometry of the pulsar as a function of time using the Precessional Rotating Vector Model \citep{Desvignes+2019,Wang+2025a}. This model relates the polarisation angle, which describes the angle of linear polarisation, to the projection of the magnetic field line direction as the pulsar beam sweeps across the line of sight \citep{Kramer+Wex2009}. By reconstructing the sky-projected polarisation emission map over the pulsar’s magnetic pole, \citet{Desvignes+2019} predict that the detectable emission will disappear by 2028. 

In this paper, we present the results of long-term timing of the relativistic binary PSR~J1906$+$0746. Section~\ref{s:Obs and Data} describes the observations and data reduction, while Section~\ref{s:Timing} presents the timing solution. The results and their implications are discussed in Section~\ref{s:Discussion}, and the conclusions are provided in Section~\ref{s:Conclusions}.

%%%%%%%%%%%%%%%%%%%% Observations and Data Reduction %%%%%%%%%%%%%%%%%%

\section{Observations and Data Reduction}
\label{s:Obs and Data}

We use data from six different radio telescopes in this paper, including archival data from Green Bank and Nançay, as well as new observations from the telescopes: Arecibo, FAST (Five-hundred-meter Aperture Spherical Telescope), Lovell, and MeerKAT.

\subsection{Observations}

Observations with the Arecibo telescope were carried out with three different backends: three (used together) Wide-band Arecibo Pulsar Processor (WAPP) filterbank machines \citep{Dowd+2000}, the Arecibo Signal Processor (ASP) machine \citep{Demorest+2004}, and the Puerto Rico Ultimate Pulsar Processing Instrument (PUPPI), which was a clone of the similarly named GUPPI\footnote{\url{https://safe.nrao.edu/wiki/bin/view/CICADA/GUPPISupportGuide}. Last time visited: 2025 December.} \citep{DuPlain+2008}. The WAPP data were converted to spectra, incoherently de-dispersed, summed, and folded at the spin period of the pulsar. The ASP data were coherently de-dispersed and folded online using the then best-known values for the dispersion measure (DM) and the spin period of the pulsar. The WAPP observations were done from 2005 June until 2009 June, while the ASP observations were done during 2006 July until 2009 August. From 2012 March to 2018 April we observed the pulsar using the PUPPI backend. These data were coherently de-dispersed at centre frequencies of 1730/1530/1380 over bandwidths of 800/700/400 MHz, respectively. Data were folded in real time with 10-s subintegrations with the best ephemeris available at the time. A calibration scan was obtained by injecting a 25 Hz noise diode into the signal path for both polarisations.

In our analysis, we also included observations taken during 2006 January until 2009 January using the Green Bank Astronomical Signal Processor (GASP) from the Green Bank Telescope (GBT) and observations from 2005 May until 2009 January with the Berkeley-Orléans-Nançay (BON) from the Nançay telescope. The GASP machine was a clone of ASP, while the BON coherent de-dispersion machine produced dedispersed and folded profiles every two minutes. The data from WAPP, ASP, GASP, and Nançay were already published in the studies from \citet{vanLeeuwen+2015} and we re-use them in this work.

Our data from the 76-m Lovell Telescope at the Jodrell Bank Observatory (JBO) were taken using two different backends: the Analogue FilterBank (AFB) and the Digital FilterBank (DFB).
The AFB data were de-dispersed and folded online at the nominal pulsar period. We used the AFB archival data from 2005 May 20 until 2009 January 08. These observations were made using a 32\,MHz bandwidth centred on 1400\,MHz. We used DFB data from that date until 2021 March 04; these observations were centred on 1400/1520\,MHz over a 512/448\,MHz bandwidth. A detailed description of the AFB backend can be found in \citet{Shemar+Lyne1996} and \citet{Hobbs+2004}, and for a description of DFB backend we refer the reader to \citet{Manchester+2013}.

We used the 64-dish MeerKAT telescope under the MeerTime project \citep{Bailes+2020} in the context of the relativistic binary (RelBin) theme \citep{Kramer+2021} to observe PSR~J1906+0746. Using the Pulsar Timing User Supplied Equipment \citep[PTUSE;][]{Bailes+2020} backend, we carried out 4 $\times$ 15\,minutes back to back monthly observations using the UHF receivers centred at a frequency of 816\,MHz with a bandwidth of 544\,MHz from 2021 August until 2023 August. The PTUSE backend acquires tied-array beam-formed voltages from the correlator-beamformer engine, part of the MeerKAT observing system, and simultaneously records coherently de-dispersed full-Stokes data both in PSRFITS \citep{Hotan+2004} format search mode filterbanks, and folded archives \citep{Bailes+2020}. The results from the searches of the search-mode data for potential detection of pulsations from the companion will be reported elsewhere. The data were coherently de-dispersed at the nominal DM and folded using the best known ephemeris. The MeerTime observations were reduced using the \texttt{MEERPIPE} pipeline \citep[see][]{Kramer+2021}, which includes radio frequency interference (RFI) excision \citep[based on a modified version of the \texttt{COASTGUARD} software;][]{Lazarus+2020} and produces cleaned archive files of varying decimation using \texttt{PAM} from the \texttt{PSRCHIVE}\footnote{\url{http://psrchive.sourceforge.net/index.shtml}} \citep{Hotan+2004} package. 

For the FAST programme, we analysed five two-hour observations taken between March and May 2022. Using the central element of the 19-beam receiver, we pointed on and away from flux calibrator PKS~2209+080 with the noise diode firing, in the FAST {\tt OnOff} mode. We next observed PSR~J1906+0746 for 2\,hrs per pointing, in {\tt Tracking} mode \citep[see also][]{Wang+2025a}. 
During the initial one minute, the noise diode continued to fire, thus tying the observation to the flux calibration source. The bandwidth of the FAST pulsar backend was 500\,MHz, around a central frequency of 1250\,MHz. The number of channels was 4096. The data were recorded in search mode with a sampling time of 49.152\,$\mu$s \citep{Nan+2011}.

A summary of all the observations used for this project is presented in Table \ref{tab:Observations}.

\begin{table*}
\setlength\tabcolsep{3pt}
  \begin{threeparttable}
    \caption{Details on the observing system and the timing dataset on PSR~J1906+0746 used in this paper.}
     \label{tab:Observations}
    \renewcommand{\arraystretch}{1.0}
     \begin{tabular}{lccccccccc}
     \hline
      \hline
Telescope & Backend & Time span & Central Freq & Bandwidth & Channel BW & \#ToAs  & EFAC & EQUAD \\
 & & & (MHz) & (MHz) & (MHz) &  &  & \\
        \midrule
Arecibo  &  WAPP  & 2005 -- 2009 &  1170,1370,1470 & 3$\times$100 & 0.195 & 22428 & 1.44(3),1.22(2),1.09(3) &  $-$4.626(9),$-$4.93(1),$-$4.637(6)\\
        & ASP & 2005 -- 2009 & 1420/1440 & 16 --32 & 4 & 218 & 2.9(3) & $-$4.61(3) \\
        & PUPPI & 2012 -- 2018 & 1730,1530,1380 & 800,700,400 & 1.56 & 153 & 0.95(6) & $-$6.2(6) \\
Green Bank & GASP & 2006 -- 2009 & 1404 & 64 & 4 & 1112 & 1.07(4) & $-$5.3(2) \\
Lovell & AFB & 2005 -- 2009 & 1402 & 64 & 1.0 & 5010 & 1.03(3) & $-$3.76(2)\\
       & DFB & 2010 -- 2022 & 1400,1520 & 512,448 &  & 147 & 0.69(7) & $-$4.12(6) \\ 
Nançay & BON & 2005 -- 2009 & 1398 & 64 & 4 & 653 & 1.6(1) & $-$4.29(6)\\
MeerKAT & PTUSE & 2021 -- 2023 & 1284,816 & 856,544 & 0.835,0.531 & 96 & 0.45(1) & -4.0(9) \\
FAST &  & 2022 Mar-May & 1250 & 500 & 0.122 & 120 & 0.9(2) & $-$4.4(8) \\

    \hline
      \hline
     \end{tabular}
    \begin{tablenotes}
      \small
      \item[*] Channel BW refers to the size of the channel bandwidth, \#ToAs refers to the number of ToAs generated per backend, and EFAC (Error factor) and EQUAD (Error quadrature) are the parameters used to model the white noise.
    \end{tablenotes}
  \end{threeparttable}
\end{table*}

\subsection{Data reduction}
 
The removal of the RFI was performed differently for each dataset. For the Jodrell AFB data, RFI was removed manually. In the case of the DFB data, a median filtering algorithm was initially applied, followed by manual removal of any remaining RFI using \texttt{PSRZAP} from the \texttt{PSRCHIVE} package. These data are not polarisation-calibrated. 

\begin{figure*}
\centering
	\includegraphics[width=15cm]{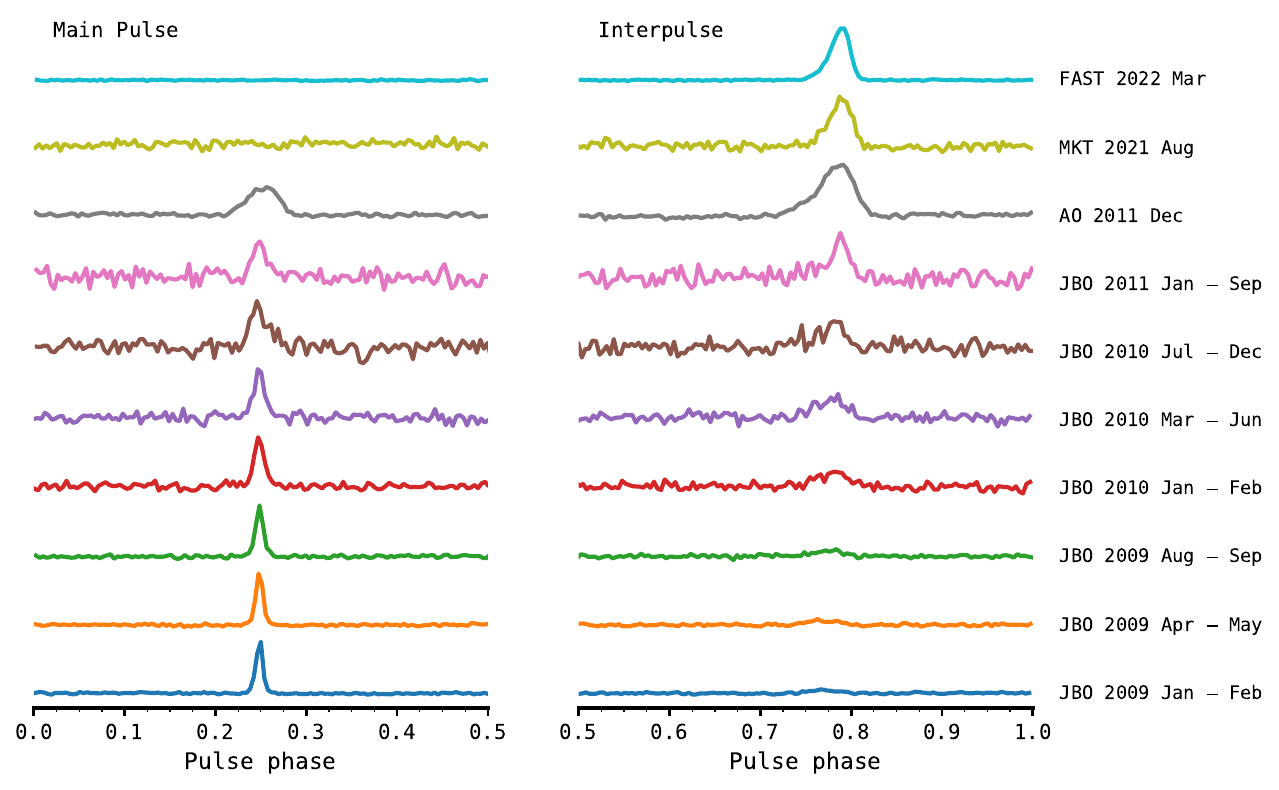}
    \caption{Integrated pulse profiles of PSR~J1906+0746 illustrating all the different profile variations seen in the DFB Jodrell (JBO), Arecibo PUPPI (AO), MeerKAT (MKT), and FAST data. The main pulse (left) and interpulse (right) are shown. For each epoch, a baseline level has been removed and the profile has been scaled such that its maximum equals unity. The dates of the observations used to obtain the profiles are indicated on the right.}
    \label{fig:pulse_profiles}
\end{figure*}

To account for all profile changes from the precession in the Jodrell DFB data, we used seven different templates covering the main period of shape evolution from 2009 January until 2011 September. We first summed the data (without including weights) that have the same pulse profile shape to obtain a mean pulse profile with higher signal-to-noise ratio (S/N; see Figure \ref{fig:pulse_profiles}). The first profile (blue) encompassed all available data from 2009 January -- February, summing a total of 15.3\,hrs; the second profile (orange) covered data from 2009 April -- May, summing a total of 20\,hrs; the third profile (green) incorporated data from 2009 August -- September, summing a total of 2.4\,hrs; the fourth profile (red) utilised data from 2010 January -- February, summing a total of 2\,hrs; the fifth profile (purple) used data from 2010 March -- June, summing a total of 2\,hrs; the sixth profile (brown) encompassed data from 2010 July -- December, summing a total of 1.7\,hrs; and for the seventh profile (pink), we included data from 2009 January -- September, summing a total of 13.2\,hrs. In order to have a consistent phase definition throughout the Jodrell data, we carefully phase-aligned those profiles manually using the profile from 2009 January (MJD 54832) as a reference, with the \texttt{PAS} routine from \texttt{PSRCHIVE}. 

The FAST data were cleaned of RFI using \texttt{PSRZAP}, then calibrated for polarization and flux using noise diode observations and the calibrator PKS 2209+080 with the \texttt{PSRCHIVE} tool \texttt{PAC}.  

For the PUPPI and MeerKAT data, we first cleaned the RFI using either \texttt{PSRZAP} or \texttt{CLFD}\footnote{\url{https://github.com/v-morello/clfd}} \citep{Morello+2019}, or both when necessary. The PUPPI data were polarisation-calibrated with \texttt{PAC} using \textit{post-facto} derived calibration solutions provided by the Arecibo Observatory. For the case of the MeerKAT observations, the data were calibrated at the telescope prior to the data being received by the PTUSE computers \citep[for a full description of the calibration method see][]{Serylak+2021}. After the data were cleaned from RFI and polarisation calibrated, they were then added in frequency, time, and polarisation (i.e. they were ``scrunched'') using \texttt{PAM}. 

The integrated profiles used to obtain the templates for extracting ToAs from the PUPPI data at the Arecibo Observatory (AO), MeerKAT, and FAST are shown in Figure~\ref{fig:pulse_profiles} in grey, olive, and light blue colours, respectively. The integrated profile of PUPPI data was obtained using the observation on 2011 December with a duration of 2.1\,hrs; the integrated profile of MeerKAT was obtained using the data from the observation on 2021 August with a duration of 30\,minutes; while the profile from the FAST data was obtained from the observation on 2022 March with a duration of 2\,hrs. We note that as the pulse profile became more stable towards the end of 2011, only one standard profile was then used for each of the PUPPI, MeerKAT, and FAST datasets. We then used \texttt{PAT} to extract the topocentric ToAs by cross-correlating the pulse profiles against the noiseless template, built by fitting von Mises functions (using the \texttt{PAAS} routine) to templates mentioned above. 

For the case of the Jodrell AFB data, we used the ToAs from the Jodrell Bank observatory archive, while the Jodrell DFB data were fully scrunched in frequency, time, and polarisation; we then obtained one ToA per observation using \texttt{PAT}. For the PUPPI data, we obtained only one ToA per observation for all the data from 2018 April 29 (MJD 58237) and earlier, since the S/N of those data were not sufficient to split the archive files into more than one subintegration. For all the PUPPI data after this date (see Table \ref{tab:Observations}), we scrunched the 2\,hr files into six subintegrations. For the MeerKAT data, we created one subintegration for every 15 minutes, and for the FAST data, we extracted ToAs every 5\,minutes. For the ToA extraction of the WAPP, ASP, GASP, and BON datasets, we refer the reader to \citet{vanLeeuwen+2015}.

%%%%%%%%%%%%%%%%%%%%%%%%%%%%%%%%%%%%%%%%%%%%%%%%%%
%%%%%%%%%%%%%%%%%%%% Timing Solutions %%%%%%%%%%%%%%%%%%

\section{Timing}
\label{s:Timing}

The measured ToAs were first converted to the Terrestrial Time (TT) realized from International Atomic Time, TT(TAI), and then referred to the Solar System Barycentre, using the JPL~DE440 planetary ephemeris \citep{Park+2021}, and fitted for different timing model parameters including celestial coordinates, spin parameters, and orbital parameters using \TEMPOTWO \citep{Hobbs+2006, Edwards+2006}. Phase connection was not possible after MJD 56583.9 (2013 October 19) indicating the presence of a large glitch, causing the pulse to be smeared in subsequent observations, rendering any pulse undetectable, until refolding with updated ephemeris was carried out. The glitch was observed around MJD 56664 $\pm$ 80 (2014 January 07), with an initial estimated size of $5.92\times 10^{-6}$ (see section \ref{s:glitch} for further details), these parameters were also included in our timing model. We then used, as our initial ephemeris, a solution obtained using only Jodrell DFB and PUPPI data, which maintains a phase coherence over our data from 2005 May (MJD 53500) to about the end of 2016 (MJD 57660).  

We used this ephemeris as our model to fit all ToAs (pre-glitch only) from the Arecibo PUPPI data to better constrain the orbital parameters before the glitch, as the Jodrell data have a lower S/N. We then fitted both datasets and used this as our starting pre-glitch ephemeris. Next, we re-folded the DFB and PUPPI data collected before the glitch with the new ephemeris and recalculated the ToAs using the standard profiles shown in Figure~\ref{fig:pulse_profiles}. We repeated this procedure for the Jodrell and Arecibo post-glitch data. Since data from different telescopes and backends are not entirely consistent due to different delays in the signal paths of each observing system, we included in our fitting procedure one ``jump'' parameter for the ToAs for each telescope and backend to account for the arbitrary time offsets between datasets. We also added a jump between the PUPPI and Jodrell data at this point. This process yielded one ephemeris for the pre-glitch data and second for the post-glitch data. We used the latter to re-fold the FAST and MeerKAT data by applying the \texttt{PAM} routine to the folded archive files and then calculated a new set of ToAs, which we added to our post-glitch solution.

For the purpose of phase-connecting the pre- and post-glitch ToAs, we used the pulse numbering feature from \TEMPOTWO. This technique allows us to track the integer number of rotations between each ToA, enabling us to maintain a continuous rotational phase across the entire 18.2\,yr dataset and achieve full phase connection. We then used this phase-connected solution to model the timing noise.

\begin{figure*}
\centering
	\includegraphics[width=12cm]{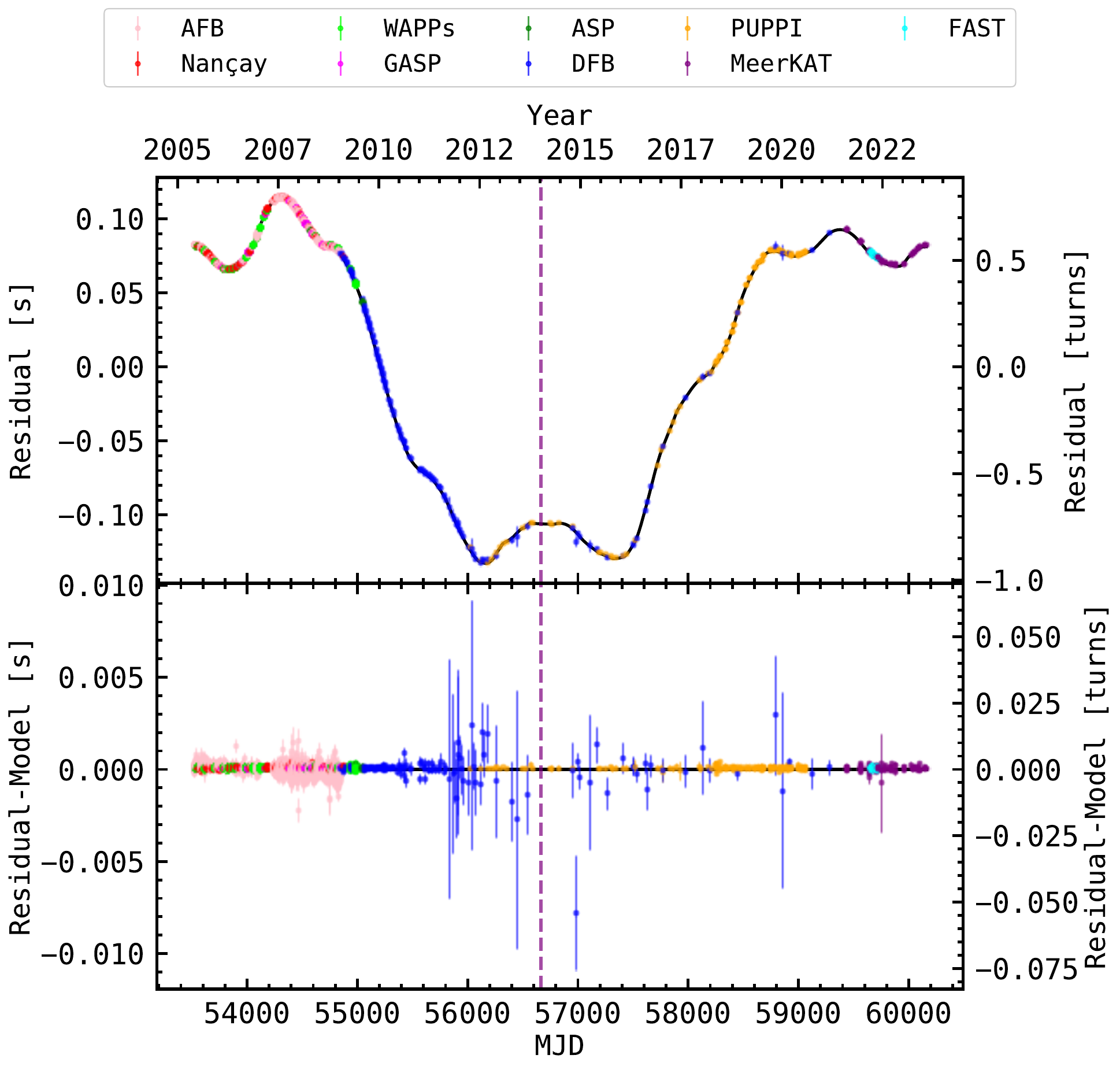}
    \caption{Phase-connected timing residuals as a function of MJD. The top panel shows the residuals we obtained as a result of the fitting procedure from \texttt{RUN\_ENTERPRISE} obtaining phase connection throughout our whole data set without removing the timing noise. The black line denotes the timing noise model fitted, while the purple dashed line denotes the glitch epoch. The bottom panel shows the result after including up to the 300th harmonic for whitening.}
    \label{fig:rednoise_residuals}
\end{figure*}

\begin{table*}
\setlength\tabcolsep{10pt}
  \begin{threeparttable}
    \caption{Timing parameters for PSR~J1906+0746, as obtained from fitting the observed ToAs with \TEMPOTWO~using the DD \citep{Damour+Deruelle1986} timing model (second column), the DDGR \citep{Taylor+Weisberg1989} model (third column), and DDGR+$\dot{x}$ (fourth column, see Section~\ref{s:x-dot}). Times are expressed in Barycentric Dynamical Time (TDB), the adopted terrestrial time standard is TT(TAI), and the Solar System ephemeris used is JPL~DE440~\citep{Park+2021}. Numbers in parentheses represent 1-$\sigma$ uncertainties in the last digit.}
    \centering
     \label{tab:timing_fitresults}
    \renewcommand{\arraystretch}{1.0}
     \begin{tabular}{l c c c} 
     \hline
      \hline
\multicolumn{4}{c}{PSR~J1906+0746} \\
\hline
Right ascension, $\alpha$ (hh:mm:ss.s; J2000) \dotfill   &  19:06:48.68(2) & 19:06:48.68(2) & 19:06:48.68(2)\\
Declination, $\delta$ (dd:mm:ss.s; J2000) \dotfill   & +07:46:27.3(4) & +07:46:27.3(4) & +07:46:27.3(4)\\
Proper motion in $\alpha$, $\mu_\alpha$ (mas\,yr$^{-1}$) \dotfill &  15(39) & 15(39) & 15(39) \\
Proper motion in $\delta$, $\mu_\delta$ (mas\,yr$^{-1}$) \dotfill & $-$12(76) & $-$12(76) & $-$12(76) \\
Spin frequency, $\nu$ (s$^{-1}$)  \dotfill 	&  6.94070335(1) & 6.94070335(1) & 6.94070335(1)\\
1st spin frequency derivative, $\dot{\nu}$ (Hz\,s$^{-1}$) \dotfill & $-$9.7579(12)$\times 10^{-13}$ & $-$9.7579(12)$\times 10^{-13}$ & $-$9.7579(12)$\times 10^{-13}$\\
2nd spin frequency derivative, $\ddot{\nu}$ (Hz\,s$^{-2}$) \dotfill & 5.8(7)$\times 10^{-24}$ & 5.8(7)$\times 10^{-24}$ & 5.8(8)$\times 10^{-24}$ \\
Epoch of period determination (MJD)  \dotfill & 56837 & 56837 & 56837 \\
Reference epoch (MJD) \dotfill & 56819.26 & 56819.26 & 56819.26 \\
Start of timing data (MJD) \dotfill & 53515.1 & 53515.1 & 53515.1 \\
End of timing data (MJD) \dotfill 	& 60159.7  & 60159.7 & 60159.7 \\
Dispersion measure, DM (pc\,cm$^{-3}$) \dotfill	& 217.843(4) & 217.843(4) & 217.843(4) \\
First derivative of dispersion measure, DM1 (cm$^{-3}$pc\,yr$^{-1}$)\dotfill & 0.0091(4) & 0.0091(4) & 0.0091(4) \\
Number of ToAs \dotfill    & 29321 & 29321 & 29321 \\
Residual rms (ms) \dotfill & 12.7 & 12.7 & 12.7 \\
\hline
\multicolumn{4}{c}{Binary Parameters}  \\
\hline
Binary model \dotfill   & DD  & DDGR & DDGR+$\dot{x}$\\
Orbital period, $P_{\rm b}$ (days) \dotfill &  0.16599304521(4) & 0.16599304525(4) & 0.16599304522(4) \\
Projected semi-major axis, $x$ (lt-s)  \dotfill &  1.4199657(8) & 1.4199546(8) & 1.419926(9) \\
Epoch of periastron, $T_0$ (MJD) \dotfill & 56836.9231353(8)  & 56836.9231358(8) & 56836.9231357(8) \\
Orbital eccentricity, $e$\dotfill & 0.0853022(4)  & 0.0852989(4) & 0.0852985(4) \\
Longitude of periastron, $\omega_0$ (deg)\dotfill & 129.2381(16) & 129.2391(16) & 129.2372(17) \\
Rate of periastron advance, $\dot{\omega}$ (deg\,yr$^{-1}$)\dotfill & 7.5841(2) & -- & --\\
First derivative of orbital period, $\dot{P_{\rm b}}$ (s\,s$^{-1}$)\dotfill & $-$5.65(2)$\times 10^{-13}$ & -- & -- \\
Einstein delay, $\gamma$ (s) \dotfill & 4.59(2)$\times 10^{-4}$ & -- & -- \\
Excess orbital period derivative, $\dot{P}_{\rm b,xs}$ (s\,s$^{-1}$)\dotfill & -- & --0.3(21)$\times 10^{-15}$ & $-$2.1(8)$\times 10^{-14}$ \\
First derivative of $x$, $\dot{x}$ (lt-s\,s$^{-1}$) \dotfill & -- & -- & $-$1.8(6)$\times 10^{-13}$ \\
Companion mass, $M_{\rm c}$ (\msun)\dotfill & -- & 1.297(5) & 1.19(3) \\
Total mass, $M_{\rm total}$ (\msun) \dotfill & -- & 2.6133(1) & 2.6133(1) \\
\hline
\multicolumn{4}{c}{Derived Parameters}  \\
\hline
Pulsar mass, $M_{\rm p}$ ($M_\odot$)\dotfill & -- & 1.316(5) & 1.42(3) \\
Rate of periastron advance, $\dot{\omega}$ (deg\,yr$^{-1}$)\dotfill & -- & 7.58(1) & 7.58(8) \\
Einstein delay, $\gamma$ (s) \dotfill & -- & 4.58(2)$\times 10^{-4}$ &  4.1(1)$\times 10^{-4}$ \\
First derivative of orbital period, $\dot{P_{\rm b}}$ (s\,s$^{-1}$)\dotfill & -- & $-$5.65(3) $\times 10^{-13}$ & $-$5.6(2) $\times 10^{-13}$ \\
Inclination angle, $i$ ($\deg$) \dotfill & -- & 44.8(2) & 50(1) \\
Galactic longitude, $l$ ($\deg$) \dotfill & 41.5979 & 41.5979 & 41.5979 \\
Galactic latitude, $b$ ($\deg$) \dotfill & 0.1475 & 0.1475 & 0.1475 \\
Spin period, $P$ (s)   \dotfill & 0.1440776171(2) & 0.1440776171(2) & 0.1440776171(2) \\ 
1st spin period derivative, $\dot{P}$ (s\,s$^{-1}$)  \dotfill & 2.0256(2)$\times 10^{-14}$ & 2.0256(2)$\times 10^{-14}$ & 2.0258(2)$\times 10^{-14}$ \\
Mass function, $f(M_{\rm p})$ (\msun)   \dotfill &  0.1115672(2)  &  0.1115646(2) & 0.111557(2) \\
Minimum companion mass, $M_{\rm c, min}$ (\msun) \dotfill &  0.8026 & 0.8026 & 0.8025 \\
Median companion mass, $M_{\rm c, med}$ (\msun)  \dotfill & 0.9758 & 0.9758 & 0.9758 \\
Surface magnetic field\tnote{${\dagger}$}, $B_0$, (G)  \dotfill &  $1.7 \times 10^{12}$ & $1.7 \times 10^{12}$ & $1.7 \times 10^{12}$ \\
Characteristic age\tnote{${\dagger}$}, $\tau_{\rm c}$ (kyr) \dotfill &  112.7 &  112.7 & 112.7 \\
\hline
\multicolumn{4}{c}{Glitch Parameters}  \\
\hline
Glitch epoch, $t_{\rm g}$ (MJD) \dotfill &  \multicolumn{3}{c}{56664.184} \\ 
Persistent step change in spin frequency, $\Delta\nu_{\rm p}$ (Hz) \dotfill &  \multicolumn{3}{c}{4.0949(9) $\times 10^{-5}$} \\
Persistent step change in spin frequency derivative, $\Delta\dot{\nu}_{\rm p}$ (Hz$^2$) \dotfill &  \multicolumn{3}{c}{--5.8(3) $\times 10^{-15}$} \\
Amplitude of the exponential recovery, $\Delta\nu_{\rm d}$ (Hz) \dotfill &  \multicolumn{3}{c}{2.15714292(2) $\times 10^{-7}$} \\
Decay time constant, $\tau_{\rm d}$ (days) \dotfill &  \multicolumn{3}{c}{100} \\
Recovery degree, $Q$ (days) \dotfill &  \multicolumn{3}{c}{0.005} \\
Fractional amplitude change in spin frequency, $\Delta \nu / \nu$ \dotfill &  \multicolumn{3}{c}{5.899(2) $\times 10^{-6}$} \\  
Fractional amplitude change in spin frequency derivative, $\Delta \dot{\nu} / \dot{\nu}$ \dotfill &  \multicolumn{3}{c}{5.9(2) $\times 10^{-3}$} \\

    \hline
      \hline
     \end{tabular}
\begin{tablenotes}
      \small
      \item[${\dagger}$] The frequency derivatives could be affected by the Galactic and centrifugal accelerations and therefore affect the values of  $B_0$ and $\tau_{\rm c}$.
\end{tablenotes}
  \end{threeparttable}
\end{table*}

\subsection{Red noise}

To model the timing noise in PSR~J1906+0746, we use \texttt{RUN\_ENTERPRISE} \citep{Keith+2022} from the \ENTERPRISE~framework \citep{Ellis+2019}. The \texttt{RUN\_ENTERPRISE} tool offers advantages for modelling the timing noise of pulsars \citep[see][for more details]{Keith+2022}. It employs a Gaussian model of the Fourier-domain power spectral density to constrain the amplitude of a harmonic series of sinusoids, i.e., the Fourier basis. We use this tool to fit a pulsar timing model that includes the spin frequency and two frequency derivatives ($\nu$, $\dot{\nu}$, and $\ddot{\nu}$); the Keplerian-binary parameters--orbital period $P_{\rm b}$, projected semi-major axis $x$, epoch of periastron $T_{0}$, and eccentricity $e$; three post-Keplerian (PK) parameters ($\dot{P}_{\rm b}$, $\gamma$, and $\dot{\omega}$); astrometric parameters (position and proper motion); clock offsets (i.e. `jumps') between all different telescope and backend systems (a total of 10 were included); a glitch model (see Section \ref{s:glitch}); and a Gaussian process model of the red noise using a power-law Fourier basis. We also fit for additional or poorly estimated white noise that might not be captured by the ToA errors using the EFAC \citep[error factor;][]{Hobbs+2006} and EQUAD \citep[quadrature factor;][]{Liu+2012} parameters (reported in Table~\ref{tab:Observations}). All these parameters were fitted simultaneously using the DD binary model \citep{Damour+Deruelle1985, Damour+Deruelle1986}, see Table \ref{tab:timing_fitresults}.

The red noise and white noise parameters are sampled using a Markov-chain Monte-Carlo (MCMC) approach, and the other parameters are marginalised over using the linearised approximation to the timing model generated by \TEMPOTWO; the sampling is performed using the \texttt{EMCEE} package\footnote{\url{https://github.com/dfm/emcee}} \citep{Foreman-Mackey+2013}.

The red noise model is parametrised by the spectral exponent $\Gamma$, and the $\log_{10}(A_{\rm red})$ of a power-law \citep[see][]{Lentati+2014,Keith+Nitu2023} defined as

\begin{equation}
\label{eq:powerlaw}
    P(f) = \frac{A^2_{\rm red}}{12\pi^2}\left(\dfrac{f}{f_{\rm yr}}\right)^{-\Gamma}f_{\rm yr}^{-3},
\end{equation}

\noindent where $f$ is the frequency in the Fourier domain and $P(f)$ is the power spectral density (PSD). The best-fit parameters we obtained are: $\Gamma = 3.45(6)$ and $\log_{10}(A_{\rm red}) = -9.17(2)$. To assess whether the relatively short-span FAST observations affect the inferred noise properties, we repeated the analysis excluding the FAST data. This yields consistent red-noise parameters ($\Gamma = 3.41(7)$, $\log_{10}(A_{\rm red}) = -9.19(3)$), indicating that the inclusion of the FAST observations does not significantly influence the results.

In the top panel of Figure \ref{fig:rednoise_residuals}, we show the timing residuals as a function of MJD obtained from the fitting procedure with \texttt{RUN\_ENTERPRISE} before removing the timing noise (weighted rms = 12.7\,ms). The black curve shows the fitted timing-noise model, while the purple dashed line indicates the glitch epoch. The bottom panel of the same figure displays the timing residuals after removing the timing noise (weighted rms = 25\,$\mu$s), modelled as a set of Fourier components using up to the 300th harmonic for whitening. This corresponds to a maximum Fourier frequency $f_{\rm max} = 300/T \approx 0.05$\,days$^{-1}$ ($\approx 16.5$\,yr$^{-1}$), where $T=6644.6$\,days is the data span. This value is set by the practical upper limit of the \TEMPOTWO\ version used in this analysis and pushes the modelling to the frequencies where the residuals begin to be dominated by white noise. Using the post-whitening rms and an effective cadence of the 591 unique observing days, we estimate a nominal white-noise level of $P_{\rm white} \approx 1.2 \times 10^{-3}\,{\rm s^3}$, while the fitted red-noise power law evaluated at $f_{\rm max}$ gives $P_{\rm red}(f_{\rm max}) \approx 7.7 \times 10^{-3}$\,s$^3$, i.e. within an order of magnitude of the nominal white-noise level. This comparison should be regarded as approximate, since the residual spectrum deviates from a single stationary power law and the effective white-noise level depends on the adopted cadence and on whether EFAC/EQUAD-adjusted ToA uncertainties are used to characterise the white component.

\begin{figure}
\centering
	\includegraphics[width=8cm]{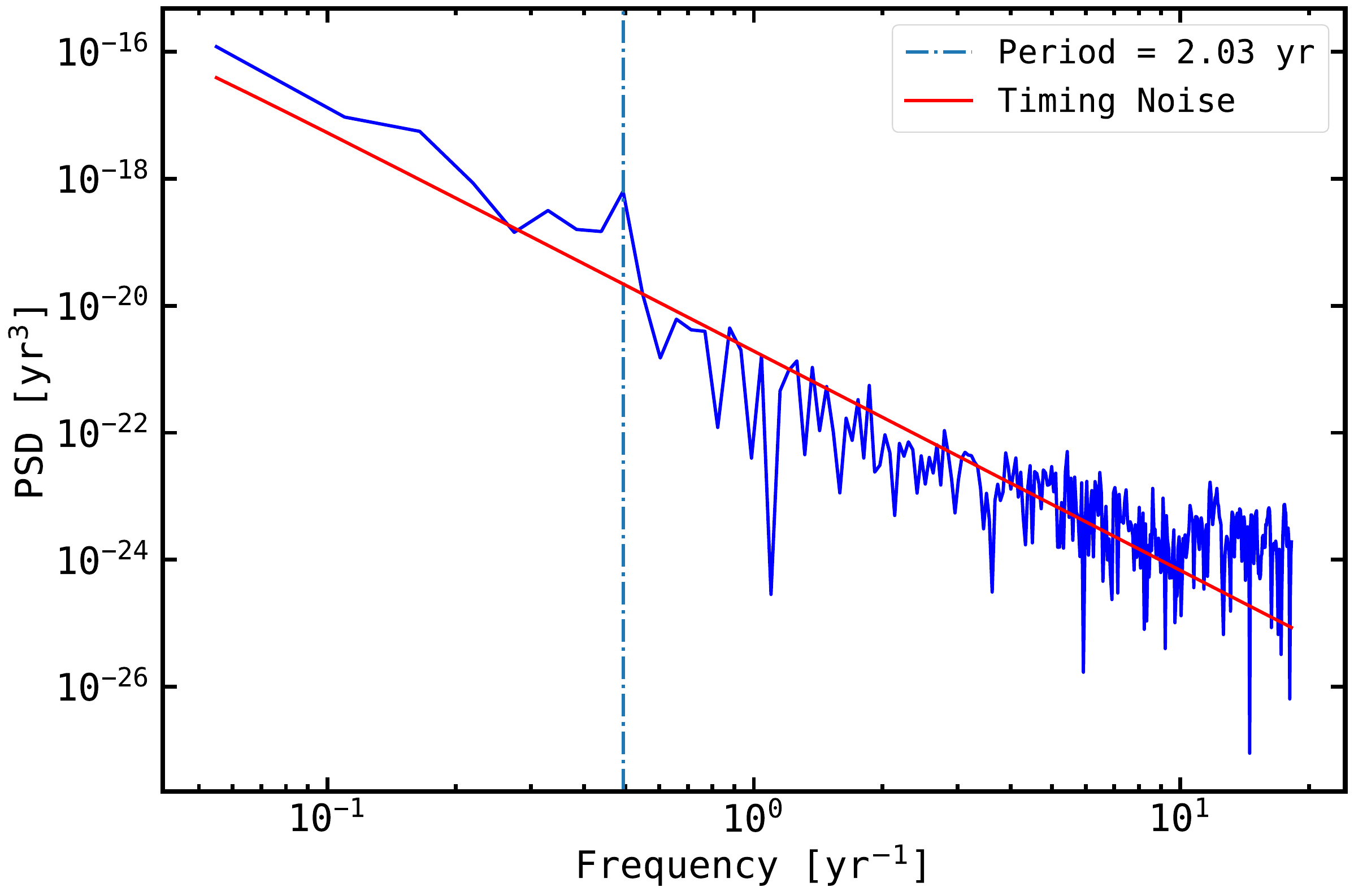}
    \caption{Power spectral density against frequency is shown by the solid blue line. The red line represents the fitted power law (Equation~\ref{eq:powerlaw}). A peak at a frequency of 0.493\,yr$^{-1}$ indicated by the vertical dash-dotted blue line, suggests evidence for a periodicity in the residuals of $\sim$2\,yr.}
    \label{fig:PSD}
\end{figure}

We show the PSD of the residuals as a function of frequency in Figure \ref{fig:PSD}, as well as the fitted power-law model (red line). The frequency range shown is limited at the low end by the time span of our observations, and at the high end by our chosen value of 0.05\,days$^{-1}$. The power spectrum resembles a broken power-law, with a steep red component transitioning to a flatter red component; this is likely because the data seem to be inconsistent with stationary power-law noise. Moreover, a peak is observed at a frequency of $\sim 0.493$\,yr$^{-1}$. This feature is seen after removing the red noise and suggests the presence of a periodicity of $\sim$2\,yr in our data (blue dash-dotted line, see Section~\ref{s:noise_discussion} for further discussion).

\subsection{Glitch}
\label{s:glitch}

A glitch is a sudden, discontinuous increase in a pulsar's spin frequency, causing pulses to arrive progressively earlier than predicted by the timing model. This change appears as an increasingly negative deviation from zero in the pulsar's timing residuals. The parameters of the glitch are typically modelled as a function of the glitch-induced phase evolution, $\phi_{\rm g}(t)$, as follows:

\begin{equation}
    \phi_{\rm g}(t) = \Delta\nu_{\rm p}(t-t_{\rm g}) + \dfrac{1}{2}\Delta\dot{\nu}_{\rm p}(t-t_{\rm g})^2 + \Delta\nu_{\rm d}\tau(1-e^{-(t-t_{\rm g})/\tau}), 
\end{equation}

\noindent where $\Delta \nu_{\rm p}$ and  $\Delta \dot{\nu}_{\rm p}$ are the persistent step changes in spin frequency and spin-frequency derivative, respectively; $t_{\rm g}$ is the glitch epoch; $\Delta \nu_{\rm d}$ is the amplitude of the exponential recovery; and $\tau_{\rm d}$ is the decay time constant. The total frequency change at the glitch is given by $\Delta \nu = \Delta \nu_{\rm p} + \Delta \nu_{\rm d}$.

\begin{figure}
\centering
	\includegraphics[width=\columnwidth]{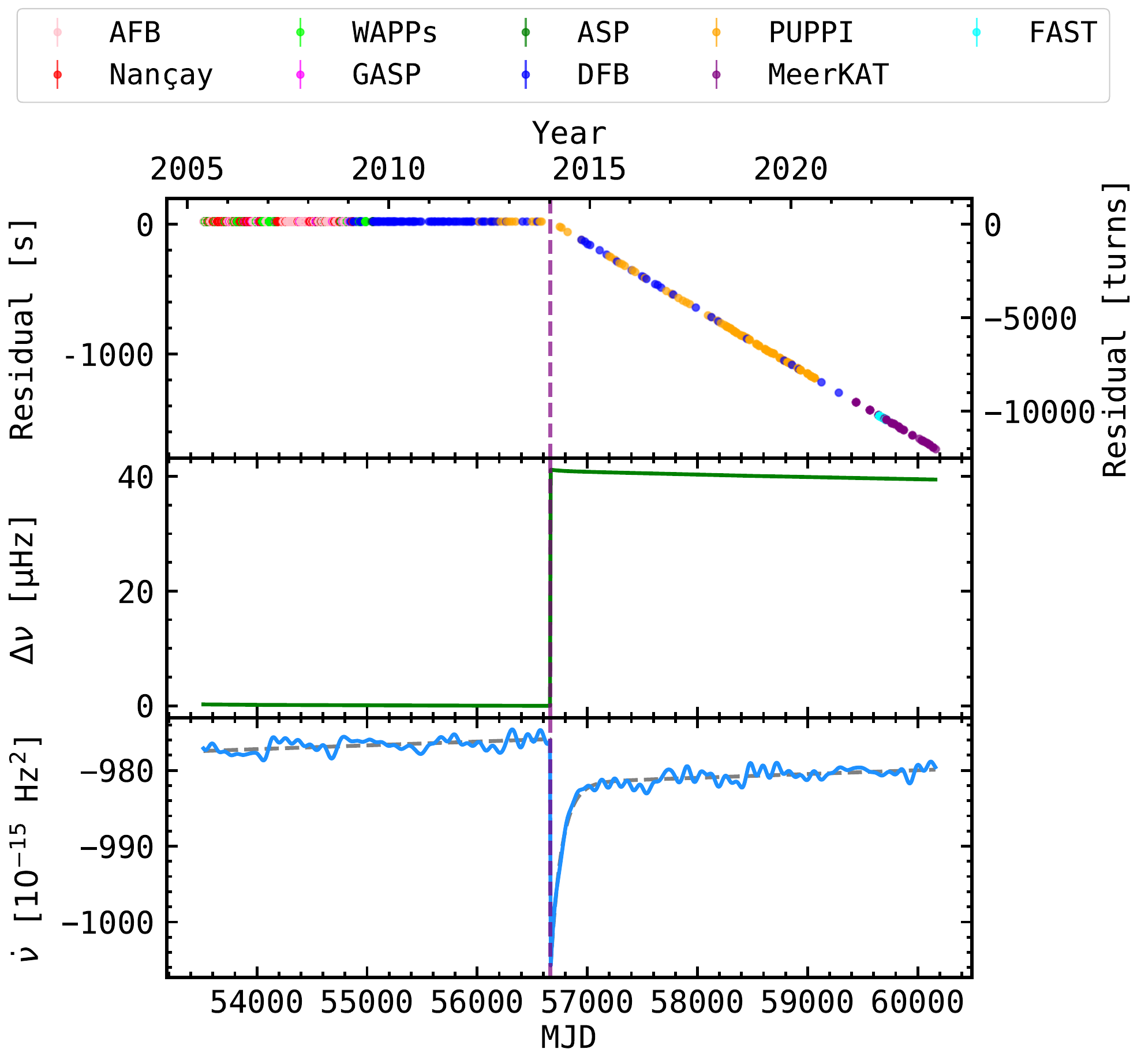}
    \caption{Timing residuals, $\Delta \nu$, and $\dot{\nu}$ as a function of time for PSR~J1906+0746 after removing the red noise including up to the 60th harmonic for whitening. The vertical purple dashed line denotes the glitch epoch.}
    \label{fig:glitch_plot}
\end{figure}

A large glitch that occurred between epochs 56583.905 (2013 October 19) and 56743.476 (2014 March 27) was identified in the data for PSR~J1906+0746. The corresponding timing residuals (after removing the timing noise) are shown in the top panel of Figure \ref{fig:glitch_plot}. In the second panel of the same figure, we compare the projected $\nu$ based on the pre-glitch data with the measured post-glitch $\nu$, revealing a step of size $\Delta \nu_{\rm p} = 4.0949(9) \times 10^{-5}$\,Hz. We also show the effect of the glitch on the spin-down rate in the third panel of the figure. The frequency first derivative,  shows a sudden jump of  $\dot{\nu} = 1.9 \times 10 ^{-14}$\,Hz$^2$ at the start of the event. The spin frequency then increases with a decay time constant of $\tau_{\rm d} \sim 100$\,days relative to the pre-glitch value, with a recovery degree $Q = \Delta \nu_{\rm d} / \Delta\nu = 0.005$. Additionally, the measured pre-glitch spin-down rate differs from the post-glitch rate, suggesting a persistent change of $\Delta \dot{\nu}_{\rm p} = -5.8(3) \times 10^{-15}$\,Hz$^{2}$.

\begin{figure}
\centering
	\includegraphics[width=8cm]{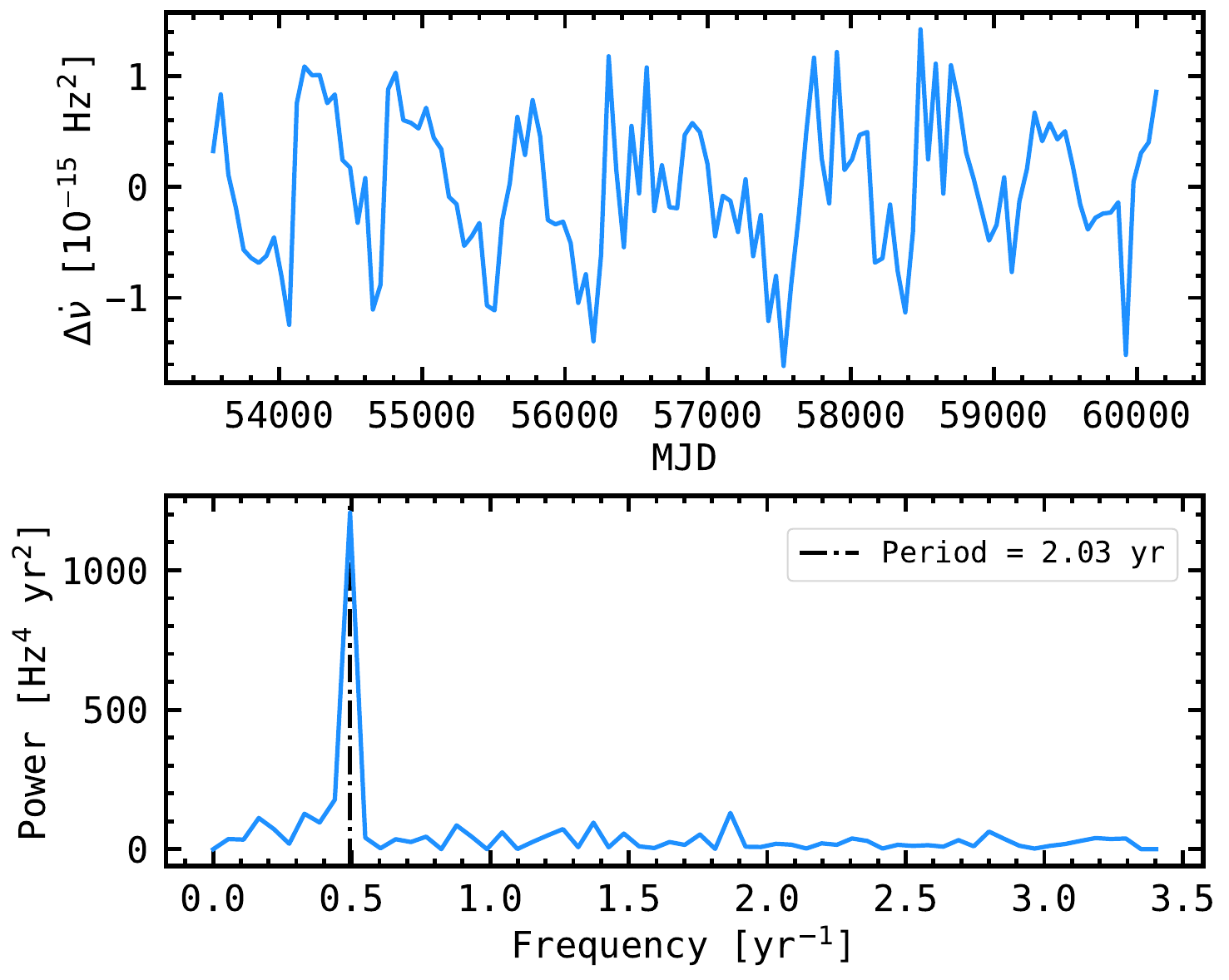}
    \caption{Top panel: evolution of $\Delta \dot{\nu}$ sampled at the Nyquist frequency. Bottom panel: Power spectral density of $\dot{\nu}$.}
    \label{fig:nudot_PSD}
\end{figure}

The fractional amplitude changes in spin frequency and its derivative are $\Delta \nu / \nu = 5.899(2) \times 10^{-6}$ and $\Delta \dot{\nu} / \dot{\nu} = 5.9(2) \times 10^{-3}$, respectively. In the top panel of Figure \ref{fig:nudot_PSD}, we show the evolution of the spin-down rate, sampled at the Nyquist frequency, while the bottom panel displays its PSD. A peak is observed at approximately 0.5\,yr$^{-1}$, which is consistent with the 2-year periodicity also evident in the PSD of the residuals. See Section \ref{s:Discussion} for further discussion.
 
\subsection{Post-Keplerian Parameters}

We modelled the relativistic binary orbit using both the theory-independent DD model and the general-relativistic DDGR model \citep{Taylor+Weisberg1989}. Three PK parameters for PSR~J1906+0746 were first detected by \citet{vanLeeuwen+2015} using a data span of only 4\,yr. In this work, we have extended the timing baseline by a factor of $\sim 4.5$. Additionally, we obtained a phase-connected solution across the entire dataset and we modelled the timing noise using the novel technique of \citet{Keith+Nitu2023}. This approach not only led to improved measurements of the three parameters but also resulted in the first 3$\sigma$ detection of the rate of change of the projected semi-major axis of the pulsar orbit, $\dot{x}$, for this system. 

\subsubsection{Advance of Periastron}

From our timing analysis, we derive a value of $\dot{\omega} = 7.5841(2)$\,deg\,yr$^{-1}$, which is approximately 2.5 times more precise than the previous measurement by \citet{vanLeeuwen+2015}, $\dot{\omega} = 7.5841(5)$\,deg\,yr$^{-1}$. Assuming the validity of general relativity (GR), this yields a total system mass of $M_{\rm total} = 2.6133(3)$\,\msun, consistent with the value derived using the DDGR model. See Table~\ref{tab:timing_fitresults}.

\subsubsection{Einstein delay}

We also measured $\gamma = 4.59(2) \times 10^{-4}$\,s, which is approximately 2.5 times more precise than the previous result from \citet{vanLeeuwen+2015}, $\gamma = 4.70(5) \times 10^{-4}$\,s, though the two measurements differ by about 2$\sigma$. This discrepancy is not statistically significant.

\subsubsection{Variation of the Orbital Period} 
\label{s:Pbdot}

The observed ToAs of the pulses are influenced by relative acceleration effects between the pulsar binary and the Solar System, which lead to modifications in the intrinsic period derivative of the pulsar, $\dot{P}_{\rm int}$. We can use the following equation to account for these effects and estimate $\dot{P}_{\rm int}$:

\begin{equation}
    \left(\dfrac{\dot{P}}{P}\right)_{\rm obs} = \left(\dfrac{\dot{P}}{P}\right)_{\rm int} - \frac{\dot{D}}{D},
\end{equation}

\noindent where $\dot{D}$ is the time derivative of the Doppler shift factor, $D$. This quantity accounts for accelerations that contribute to the observed period derivative, $\dot{P}_{\rm obs}$. Specifically, $\dot{D}$ includes: (1) the line-of-sight acceleration due to the difference in the Galactic gravitational potential acting on the pulsar and the Solar System, denoted as $a_{\rm Gal}$; and (2) the apparent acceleration caused by the transverse Doppler effect (also known as the Shklovskii effect; \citealt{Shklovskii1970}), which arises from the pulsar's proper motion $\mu$ and is given by $a_{\rm PM} = \mu^2 d$, where $d$ is the distance to the pulsar. Therefore,

\begin{equation}
\label{eq:Doppler}
    \frac{\dot{D}}{D} \equiv - \dfrac{a_{\rm Gal} + a_{\rm PM}}{c}.
\end{equation}

\noindent At the same time, the Galactic acceleration is given by the sum of the contributions from both the Galactic rotation, $a_{\rm Gal,rot}$, and the vertical acceleration towards the Galactic disc, $a_{\rm Gal,disc}$ \citep{Damour+Taylor1991, Nice+Taylor1995, Lazaridis+2009}. The total Galactic acceleration is therefore given by 

\begin{equation}
    a_{\rm Gal,rot} = -\frac{\Theta^2_0}{R_0}\left(\cos l + \frac{\xi}{\xi^2+\sin^2 l}\right) \cos b,
\end{equation}

\begin{equation}
    a_{\rm Gal,disc} = - \left[2.27z_{\rm kpc} + 3.68(1-\exp^{-4.3z_{\rm kpc}})\right]\abs{\sin b},
\end{equation}

\noindent where $z_{\rm kpc} \equiv \abs{d \sin b}$ is the Galactic height in kpc, and $\xi \equiv (d/R_{\odot})\cos b - \cos l$. We adopt a Galactic rotation velocity of $\Theta_0 = 240.5(4)$\,km\,s$^{-1}$ and a distance to the Galactic centre of $R_0 = 8.275(34)$\,kpc \citep{Guo+2021,Grav+2021}. Given the low Galactic latitude of PSR~J1906+0746 ($b = 0.1475^\circ$), the vertical acceleration component is nearly perpendicular to the line of sight, making the contribution from Galactic rotation, $a_{\rm Gal,rot}$, the dominant term. 

Using the H\RNum{1} distance estimate of $7.4^{+2.5}_{-1.4}$\,kpc from \citet{vanLeeuwen+2015}, along with the pulsar's Galactic coordinates and proper motion values from Table~\ref{tab:timing_fitresults}, we derive an intrinsic period derivative of $\dot{P}_{\rm int} = 2.02(3) \times 10^{-14}$\,s\,s$^{-1}$, which is very similar to the uncorrected value. However, we note that $\dot{P}_{\rm int}$ is strongly correlated with the strength of the timing noise, and some red noise power may be absorbed into this value \citep{Parthasarathy+2019}. Using this measurement and the relations in \citet{Lorimer+Kramer2004}, we estimate the surface magnetic field to be $B_{\rm s} \simeq 1.7 \times 10^{12}$\,G, the characteristic age $\tau_{\rm c} \simeq 112.8$\,kyr, and the rotational energy loss rate $\dot{E} \simeq 2.7 \times 10^{35}$\,erg\,s$^{-1}$.

The observed orbital period derivative, $\dot{P}_{\rm b,obs} = -5.65(2)\times 10^{-13}$\,s\,s$^{-1}$, which is $\sim 15$ times more precise than the estimate by \citet{vanLeeuwen+2015}, is also affected by the same acceleration effects, so that we have 

\begin{equation}
\label{eq:Pbdot}
    \left(\dfrac{\dot{P}_{\rm b}}{P_{\rm b}}\right)_{\rm obs} = \left(\dfrac{\dot{P}_{\rm b}}{P_{\rm b}}\right)_{\rm GW} + \left(\dfrac{\dot{P}_{\rm b}}{P_{\rm b}}\right)_{\rm Gal} + \left(\dfrac{\dot{P}_{\rm b}}{P_{\rm b}}\right)_{\rm PM} + \left(\dfrac{\dot{P}_{\rm b}}{P_{\rm b}}\right)_{\dot{m}}.
\end{equation}

We calculate the Galactic contribution as $\dot{P}_{\rm b,Gal} = (a_{\rm Gal}P_{\rm b}) /c = -0.12(6)\times 10^{-13}$\,s\,s$^{-1}$ and the Shklovskii contribution as $\dot{P}_{\rm b,PM} = (a_{\rm PM}P_{\rm b}) /c  = 0.95(2.9) \times 10^{-13}$\,s\,s$^{-1}$. The final term in Equation~\ref{eq:Pbdot} arises from mass loss in the binary system  and is given by \citet{Damour+Taylor1991}: 

\begin{equation}
\label{eq:Pbdot_m}
    \left(\dfrac{\dot{P}_{\rm b}}{P_{\rm b}}\right)_{\dot{m}} = \dfrac{8\pi^2I}{c^2M_{\rm total}}\dfrac{\dot{P}}{P^3},
\end{equation}

\noindent where $I \simeq 10^{45}$\,g\,cm$^{2}$ is the moment of inertia of the pulsar. Using this expression, we estimate the orbital period derivative due to mass loss to be $\dot{P}_{\rm b,\dot{m}} \sim 1.6 \times 10^{-15}$\,s\,s$^{-1}$, which is negligible compared to the observed value $\dot{P}_{\rm b,obs}$. We then use the $\dot{P}_{\rm b,\mathrm{obs}}$ value obtained from the DD binary model, and subtracting the Galactic and Shklovskii contributions, we obtain an intrinsic orbital period derivative of $\dot{P}_{\rm b,GW} = 0.7(3) \times 10^{-12}$\,s\,s$^{-1}$. However, the fitted proper motion has large uncertainties and is consistent with zero. Nonetheless, we can place an upper limit on the centrifugal acceleration due to the transverse Doppler contribution and the system's proper motion. Assuming that the intrinsic orbital decay matches the general relativistic prediction from the DDGR model,
$\dot{P}_{\rm b,GR} = -5.65(3) \times 10^{-13}$\,s\,s$^{-1}$,
and combining it with the $\dot{P}_{\rm b,Gal}$ and $\dot{P}_{\rm b,obs}$ terms, we find:

\begin{equation}
    \dot{P}_{\rm b,PM} = \dfrac{\mu^2 dP_{\rm b}}{c} < 0.2\times 10^{-13}\,{\rm s}\,{\rm s}^{-1}
\end{equation}

\noindent (at the 95\% confidence level, CL). This corresponds to an upper limit on the total proper motion of $\mu < 10$\,mas\,yr$^{-1}$ or a transverse velocity $v_{\rm T} = \mu d < 600$\,km\,s$^{-1}$ (95\% CL), implying that the proper motion value from our timing solution $\mu = 19(55)$\,mas\,yr$^{-1}$ is consistent with the value from the GR prediction. Our results thus imply that the $\dot{P}_{\rm b}$ value is also consistent with that predicted by GR.

\subsubsection{Variation of the projected semi-major axis}
\label{s:x-dot}

We now fit for the variation of the projected semi-major axis ($\dot{x}$) of PSR~J1906+0746. The motivation for this is discussed in the following section. In the DDGR binary model, we obtain a mildly significant 3$\sigma$ measurement of $\dot{x}_{\rm obs} = -1.8(6)\times 10^{-13}$\,lt-s\,s$^{-1}$ (see Table~\ref{tab:timing_fitresults}). This can arise due to a number of geometric and physical effects, and can be written as follows \citep{Lorimer+Kramer2004}:

\begin{equation}
     \left(\dfrac{\dot{x}}{x}\right)_{\rm obs} = \left(\dfrac{\dot{x}}{x}\right)_{\rm GW} + \left(\dfrac{\dot{x}}{x}\right)_{\rm PM} + \dfrac{\dev \epsilon_A}{\dev t} - \dfrac{\dot{D}}{D} + \left(\dfrac{\dot{x}}{x}\right)_{\dot{m}} + \left(\dfrac{\dot{x}}{x}\right)_{\rm SO}^{\rm p} + \left(\dfrac{\dot{x}}{x}\right)_{\rm SO}^{\rm c}. 
\end{equation}

\noindent We now describe each term and estimate its magnitude. The first term arises from the decrease in the size of the orbit due to gravitational wave emission and is given by:

\begin{equation}
\label{eq: xdot_GW}
    \left(\dfrac{\dot{x}}{x}\right)_{\rm GW} = \frac{2}{3}\frac{\dot{P}_{\rm b, GW}}{P_{\rm b}} = -2.6 \times 10^{-17}\,{\rm s}^{-1},
\end{equation}

\noindent with $\dot{P}_{\rm b, GW}$ is as predicted by GR. We then obtain that $\dot{x}_{\rm GW}= -3.7 \times 10^{-17}$\,lt-s\,s$^{-1}$. This is four orders of magnitude smaller than $\dot{x}_{\rm obs}$.  

The second term arises from the effects of the system's proper motion. The maximum contribution to $\dot{x}_{\rm obs}$ due to proper motion is given by \citep{Arzoumanian+1996, Kopeikin1996}:

\begin{equation}
    \left(\dfrac{\dot{x}}{x}\right)_{\rm PM} \leq 1.54 \times 10^{-16} \cot i \left( \frac{\mu}{\rm mas\,yr^{-1}}\right),
\end{equation}

\noindent we then have $\dot{x}_{\rm PM} = 3.5 \times 10^{-15}$\,lt-s\,s$^{-1}$. This value is two orders of magnitude smaller than the observed $\dot{x}_{\rm obs}$. 

The third term is caused by a change in the aberration due to geodetic precession. The latter is giving by \citep{Barker+OConnell1975}: 

\begin{equation}
    \Omega_{\rm geod} = \left( \frac{2\pi}{P_{\rm b}} \right)^{5/3} {\rm T}_{\odot}^{2/3} \frac{1}{1-e^2} \frac{M_{\rm c}(4M_{\rm p} + 3M_{\rm c})}{2M_{\rm total}^{4/3}}.
\end{equation}

\noindent Assuming the mass values as obtained from the DDGR+$\dot{x}$ model (see Table~\ref{tab:timing_fitresults}) and the Keplerian parameters of the system, we obtain $\Omega_{\rm geod} = 2.04$\,$\deg$\,yr$^{-1}$. The change in the aberration is then proportional to the $\Omega_{\rm geod}$ \citep{Damour+Taylor1992}:

\begin{equation}
    \dfrac{\dev \epsilon_{A}}{\dev t} = \frac{P}{P_{\rm b}} \frac{\Omega_{\rm geod}}{\sqrt{1-e^2}} \frac{\cot \lambda \sin 2\eta + \cot i \cos \eta}{\sin \lambda},
\end{equation} 

\noindent where $\eta$ and $\lambda$ are the two polar angles that give the orientation of the precessing spin axis. Using the geometric parameters obtained from the rotation vector model by \citet{Desvignes+2019}, specifically the magnetic inclination angle $\alpha = 99.4^\circ$ and the spin-orbit misalignment angle $\delta = 104^\circ$, we find an impact parameter of $\beta = -14.6^\circ$ at the reference epoch $T_0$.  Together with the inclination angle from the DDGR+$\dot{x}$ timing model, $i = 50(2)\,\deg$, and applying Equation 8 from \citet{Kramer+Wex2009} to estimate $\eta(t)$, we find that the contribution to $\dot{x}_{\rm obs}$ is approximately $4.8 \times 10^{-14}$\,lt-s\,s$^{-1}$, accounting for about 27\% of the total observed value.

The fourth term arises from the variation of the Doppler shift due to the system’s proper motion on the plane of the sky, as given by Equation~\ref{eq:Doppler}. Using the values discussed in Section~\ref{s:Pbdot}, we find that the $\dot{D}/D$ contribution to $\dot{x}_{\rm obs}$ is $8.3 \times 10^{-18}$\,lt-s\,s$^{-1}$, which is more than four orders of magnitude smaller than the observed value and thus negligible. 

The fifth term accounts for changes in the projected semi-major axis due to mass loss from the system. This effect can be derived using Equation~\ref{eq:Pbdot_m} and translated into a change in $x$ via Equation~\ref{eq: xdot_GW}. We obtain $\dot{x}_{\dot{m}} = 7.5 \times 10^{-17}$\,lt-s\,s$^{-1}$, which is also negligible.

This leaves the last two terms in the $\dot{x}_{\rm obs}$ contributions as the major plausible explanation for the observed value. These terms arise from spin-orbit coupling due to the rapid rotation of the pulsar ($\dot{x}_{\rm SO}^{\rm p}$) and its companion ($\dot{x}_{\rm SO}^{\rm c}$). Each of these contributions includes two components: the classical spin-orbit coupling ($\dot{x}_{\rm QM}$) and the relativistic spin-orbit coupling, also known as the Lense-Thirring effect ($\dot{x}_{\rm LT}$), which is caused by the rotation of the pulsar or the companion \citep{Lense+Thirring1918, Barker+OConnell1975, Damour+Taylor1992}. For main-sequence stars, $\dot{x}_{\rm QM} \gg \dot{x}_{\rm LT}$; for neutron stars (NSs), the opposite holds; and for WDs, the two contributions are typically comparable.

The classical quadrupole-induced contribution is given by \citep{Smarr+Blandford1976, Lai+1995, Wex+1998}:

\begin{equation}    
    \dot{x}_{\rm QM} = x\left( \frac{2\pi}{P_{\rm b}} \right) Q \cot i \sin \delta_{\rm A} \cos \delta_{\rm A} \sin \Phi_{\rm A}^{0},
\end{equation}

\noindent where

\begin{equation}
    Q = \frac{k_2R^2_{\rm A}\hat{\Omega}^2_{\rm A}}{a^2(1-e^2)^2} \,\,\, {\rm and} \,\,\,  \hat{\Omega}_{\rm A} \equiv \frac{\Omega_{\rm A}}{(GM_{\rm A}/R^3_{\rm A})^{\frac{1}{2}}},
\end{equation}

\noindent with $\Omega_{\rm A} = 2\pi/P_{\rm A}$ is the angular spin frequency of component A. The angles $\delta_{\rm A}$ and $\Phi_{\rm A}^{0}$ define the orientation of the spin axis of component A with respect to the orbital plane. The parameters $M_{\rm A}$, $P_{\rm A}$, and $R_{\rm A}$ denote the mass, spin period, and radius of component A, respectively; $k_2$ is the apsidal motion constant; and $a$ is the orbital separation, computed using Equation 3 from \citet{Ridolfi+2019}.

The relativistic contribution from the Lense-Thirring effect is given by \citep{Damour+Taylor1992}:

\begin{equation}
    \dot{x}_{\rm LT} \simeq -x \frac{GS_{\rm A}}{c^2a^3(1-e^2)^{3/2}} \left( 2+ \frac{3M_{\rm B}}{2M_{\rm A}} \right) \cot i \sin \delta_{\rm A} \sin \Phi^0_{\rm A},
\end{equation}

\noindent where $S_{\rm A} = I_{\rm A} \Omega_{\rm A}$ is the spin angular momentum of component A, with $I_{\rm A}$ its moment of inertia, and $M_{\rm B}$ the mass of the companion (component B). 

We now estimate the expected contributions from both the pulsar and its companion. For the pulsar, assuming $R_{\rm p} = 12.5$\,km, $k_2 = 0.13$, and $\Omega_{\rm p} = 2\pi/P = 43.60$\,s$^{-1}$, we obtain a dimensionless spin parameter $\hat{\Omega}_{\rm p} \sim 3 \times 10^{-3}$. This gives a quadrupole moment $Q_{\rm p} \sim 6.2 \times 10^{-17}$, and consequently, we find:

\begin{equation}
    \dot{x}_{\rm QM}^{\rm p} \sim 10^{-19} \sin \delta_{\rm p} \cos \delta_{\rm p} \sin \Phi_{\rm p}^{0}\,\,\text{lt-s}\,\text{s}^{-1}, 
\end{equation}

\noindent and hence irrelevant. In this case, the angular momentum of $S_{\rm p} \sim 4.4 \times 10^{39}$\,kg\,m$^{2}$\,s$^{-1}$, leading to:

\begin{equation}
    \dot{x}_{\rm LT}^{\rm p} \sim -7 \times 10^{-15} \sin \delta_{\rm p} \sin \Phi^0_{\rm p} \,\,\text{lt-s}\,\text{s}^{-1},
\end{equation}

\noindent which is $\sim 2$ orders of magnitude smaller than the observed value.

\noindent If the companion is another NS, the $\dot{x}_{\rm LT}^{\rm c, NS}$ contribution to $\dot{x}_{\rm obs}$ is of the same order if the spin period is of the order of 10\,ms. 

Alternatively, if the companion is a WD with a mass of $1.19\,M_{\odot}$, according to \citet{Boshkayev+2017}, we estimate a radius $R_{\rm c, WD} \simeq 4000$\,km. Using the equation-of-state and composition independent I-Love-Q relations \citep{Boshkayev+2017} for the WD, we estimate $k_2 = 0.1$. Assuming a WD with a spin period $P_{\rm c,WD} = 4$\,minutes, since the value of $\dot{x}_{\rm obs }$ is similar to that of \citet[][$\dot{x}_{\rm obs} = 1.7(3) \times 10^{-13}$\,lt-s\,s$^{-1}$]{Venkatraman+2020}, we then have $\hat{\Omega}_{\rm c,WD} \sim 0.03$, and $Q_{\rm c,WD} \sim 4.4 \times 10^{-10}$. This leads to:

\begin{equation}
    \dot{x}_{\rm QM}^{\rm c, WD} \sim 2.3 \times 10^{-13} \sin \delta_{\rm c} \cos  \Phi_{\rm c}^{0}\,\,\text{lt-s\,s}^{-1},
\end{equation}

Assuming a moment of inertia $I_{\rm c,WD} \simeq 5.5 \times 10^{49}$\,g\,cm$^2$ \citep[also from][]{Boshkayev+2017}, we estimate $S_{\rm c,WD} = 1.7 \times 10^{41}$\,kg\,m$^2$\,s$^{-1}$, and therefore:

\begin{equation}
   \dot{x}_{\rm LT}^{\rm c, WD} \sim - 3.1 \times 10^{-13} \sin \delta_{\rm c,WD} \sin \Phi^0_{\rm c,WD}\,\,\text{lt-s\,s}^{-1}. 
\end{equation}

Thus, if the companion is another NS, it must have a spin period of around 10\,ms to produce a comparable contribution. However, we cannot exclude the possibility that the companion is a fast-spinning WD due to the non-zero value of $\dot{x}_{\rm obs}$ and the changes in inferred masses that arise from the correlation between $\dot{x}$ and $\gamma$ \citep[see e.g.][]{Ridolfi+2019, Venkatraman+2020}. Additionally, the value of $\gamma$ has become significantly more uncertain, which is expected from this correlation. As already discussed in \citet{Lorimer+2006} and \citet{vanLeeuwen+2015}, the expected age of the WD companion is $\sim 1\,\mathrm{Myr}$. For a white dwarf (WD) with this age and a mass of $\sim 1.2\,M_{\odot}$, evolutionary models predict absolute magnitudes of $M_V \approx 10.6$, $M_J \approx 11.4$, and $M_K \approx 11.6$ \citep{Salaris+2022}. Assuming a distance of $d_{\mathrm{H,I}} = 7.4\,\mathrm{kpc}$, the corresponding distance modulus is $(m - M) \approx 14.3$, where $M_V$, $M_J$, and $M_K$ are the absolute magnitudes in the $V$, $J$, and $K$ bands, respectively, $m$ and $M$ denote apparent and absolute magnitude. However, extinction along the line of sight is expected to be severe since PSR~J1906$+$0746 may lie behind the molecular cloud CHIMPS~4497 \citep{Rigby+2019}, in which case the optical and near-infrared reddening would be extreme. Dust maps indicate extinctions of $A_V = 33.9$--$39.4$, $A_J = 9.0$--$11.5$, and $A_K = 3.8$--$4.7$\,mag \citep{Schlegel+1998, Schlafly+Finkelbeiner2011}. These values imply apparent optical magnitudes of $V \approx 58.8$--$64.3$, too faint for a realistic detection with the current optical facilities like the Hubble Space Telescope (HST). At near-infrared wavelengths the extinction is significantly reduced; in particular, the expected magnitude in the James Webb Space Telescope (JWST) Near Infrared Camera (NIRCam) F356W band is $m_{\mathrm{F356W}} \approx 29.7$--$30.6$. While such a source could in principle be accessible to very deep JWST observations, the extremely low Galactic latitude of PSR~J1906$+$0746 ($b = 0.15^{\circ}$) implies substantial crowding, which would significantly affect detectability.

\subsection{Individual mass measurements}

\begin{figure}
\centering
	\includegraphics[width=\columnwidth]{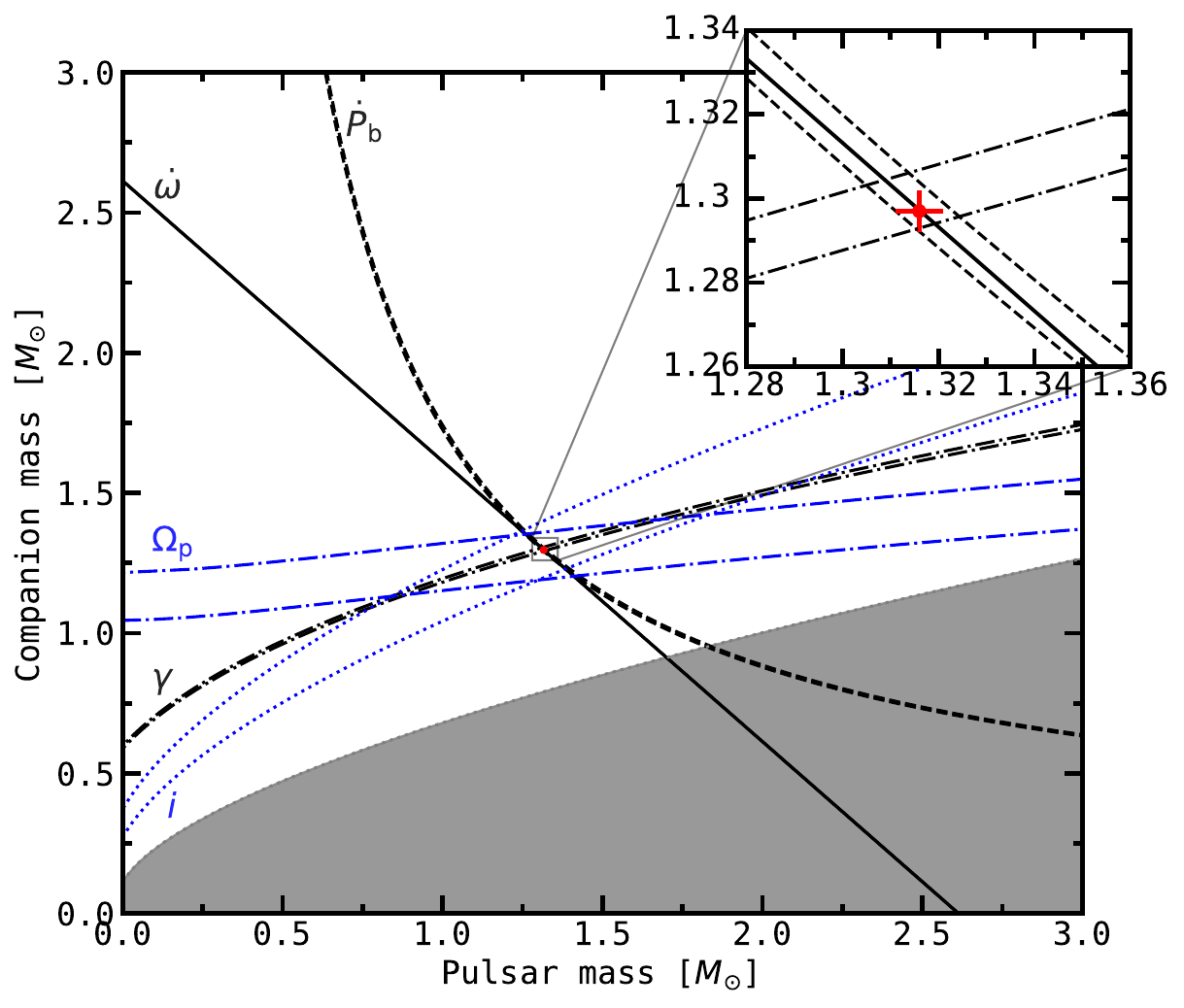}
    \caption{Mass-mass diagram of PSR~J1906+0746. The solid lines represent the values of $\dot{\omega}$, the dashed lines shows the values of $\dot{P}_{\rm b}$, and the dash-dotted lines display the values of $\gamma$ as measured by the DD binary model in Table \ref{tab:timing_fitresults}. The red dot indicates the best-fit value for $M_{\rm p}$ and $M_{\rm c}$ obtained from the DDGR model. Mass values in the grey region are excluded by the mass function and the condition $i \leq 90^{\circ}$. The dotted and dashed-dotted blue lines display the measured values from $\Omega_{\rm p}$ and $i$  by \citet{Desvignes+2019}.}
    \label{fig:m-m_diagram}
\end{figure}

We use the intrinsic values measured for the advance of the periastron $\dot{\omega}$, the Einstein delay $\gamma$, and the orbital decay $\dot{P}_{\rm b}$ to constraint the masses of the components of the binary system, the pulsar and its companion under the assumption of GR. The results are shown in the mass-mass diagram in Figure~\ref{fig:m-m_diagram}. In the same figure, we also show the measurements from \citealt{Desvignes+2019}, obtained using an independent analysis, of two additional PK parameters: the spin precession rate, $\Omega_{\rm p} = 2.17(11)\deg$\,yr$^{-1}$, and the inclination angle, $i = 45(3)^{\circ}$, represented by the dotted and dashed-dotted blue lines, respectively. All PK parameters agree in a particular region which coincides with the masses derived from the DDGR solution: $M_{\rm p} = 1.319(5)$\,\msun~ and $M_{\rm c} = 1.297(5)$\,\msun, shown in the diagram as a red cross. This agreement indicates that GR predicts the behaviour of the binary system to better than 1\% precision. These values are within 2-$\sigma$ from the previous estimates from \citet{vanLeeuwen+2015}, but the uncertainties in our measurements are a factor of $\sim 2$ smaller.

On the other hand, given the correlation of $\dot{x}$ with $\gamma$ \citep{Ridolfi+2019}, the inclusion of $\dot{x}$ in the fitting shifts the estimate of $\gamma$ and thus of the individual masses by approximately $3.5\sigma$. In this case, the companion mass is $1.19(5) \, M_{\odot}$ and the pulsar mass is 
$1.42(5) \, M_{\odot}$ shown in Figure~\ref{fig:m-m_diagram_xdot}
as the red cross.

\section{Discussion}
\label{s:Discussion}

\subsection{Glitch and Timing Noise}
\label{s:noise_discussion}

We have presented the results of over 18\,yr-timing analysis of PSR~J1906+0746, combining data from six different telescopes: Arecibo, GBT, Jodrell, Nançay, MeerKAT, and FAST using the novel technique for modelling the timing red noise seen in the pulsar as a Fourier-basis Gaussian process, as described in \citet{Keith+Nitu2023}. 

In addition to the red noise, we observed a glitch with a size of $\Delta \nu / \nu \sim 5.899(2) \times 10^{-6}$. This magnitude is comparable to the glitches observed in the Vela pulsar, which have a relative size of $\Delta \nu / \nu \sim 10^{-6}$ \citep{Yu+2013}. However, the occurrence of glitches differs, the Vela pulsar experiences glitches on a quasi-regular basis \citep{Wang+2000}. Similar glitch sizes have been reported in other pulsars of similar characteristic ages, such as PSR~J1052-5954 \citep[$\tau \sim $144\,kyr;][]{Weltevrede+2010} in which only one glitch has also been observed. Due to the sparsely sampled data around the epoch of the glitch, we were not able to determine the glitch epoch to better than $\sim 80$\,days by setting the glitch epoch midway between the observation just before and just after the glitch. Moreover, we note the fact that the glitch epoch (MJD 56664) is close to the estimated date of the polarisation angle flip (MJD~56800; \citealt{Desvignes+2019}) is a coincidence, as the polarisation change depends on the viewing geometry, whereas the glitch is an intrinsic pulsar phenomenon. The glitch shows clear relaxation, well described by an exponential decay model with a timescale of $\tau_{\rm d} \sim 100$\,days and a recovery fraction of $Q = 0.005$, resulting in a persistent offset in the value of $\dot \nu$ over time. 

The suggested periodicity of $\sim 2$\,yr found both in the PSD of the residuals and in the $\dot{\nu}$ variations, appears to be evidence of a  periodic signal present in the data. Under the hypothetical interpretation of a planetary companion, we used an extended version of \texttt{ENTERPRISE} \citep{Keith+2022} to fit for parameters including the orbital period, companion mass, white noise, and the noise properties, among others, using the timing solution of the pulsar \citep[see][for further detalis]{Nitu+2022}. The fit finds a planet-like influence with an orbital period of 736\,days, a projected mass of 4\,\mearth, and a circular orbit ($e = 2 \times 10^{-5}$). \citet{Lyne+2010} characterised several pulsars showing quasi-periodic behaviours both in $\nu$ and $\dot \nu$ and concluded that they were correlated with pulse profile changes suggesting that these were caused by changes in the pulsars’ magnetosphere. We defer a study of pulse profile variability and further investigation of the periodicity’s origin to future work.

If a planetary companion exists, PSR~J1906+0746 would be part of a triple system, and the planet would have had to survive the two supernova (SN) explosions, in the case that the companion is another NS, or one explosion in the less likely scenario if the companion is a massive WD \citep{Lorimer+2006}. 

\subsection{Nature of the companion}

Our measurements of the three PK parameters are in agreement with the previous values reported by \citet{vanLeeuwen+2015}. The uncertainties of both $\dot \omega$, and $\gamma$ are $\sim 2.5$ times smaller. Notably, the most significant improvement is in the measurement of the first derivative of the orbital period $\dot{P}_{\rm b}$, whose uncertainty is $\sim 7$ times smaller. Furthermore, the estimates of the masses of the pulsar and its companion obtained using the DDGR binary model are within 2$\sigma$ of the previous estimations, but the uncertainties are $\sim 2$ times smaller. Additionally, we report the first 3$\sigma$ detection of $\dot{x}$ for this system. This value includes a small contribution from geodetic precession of the pulsar arising from the change in the aberration of the system \citep{Damour+Taylor1992} and the largest contribution comes from the spin-orbit coupling due to the rotation of the companion. According to our calculations, the spin period of the companion needs to be of the order of 10\,ms for it to be another NS, meaning that the companion was recycled first, and the currently observed 144\,ms pulsar is the young NS, formed in the second SN. As discussed in \citealt{vanLeeuwen+2015}, the current eccentricity ($e = 0.0853022(4)$) reflects the state of the orbit post-SN configuration. In this case, the first NS have to been spun up by accretion from its companion, meaning that these type of pulsars have far lower magnetic fields than the general pulsar population and thus show very stable rotation and evolve slowly. This stability and longevity mean such recycled pulsars are observable for much longer periods compared to the fast-evolving young pulsars \citep{Srinivasan+vandenHeuvel1982}. Searches for pulsed emission were carried out using WAPP and Spigot data \citep{vanLeeuwen+2015}, but no pulsar signal was detected. The presumed recycled pulsar may nevertheless become visible on a geodetic precession timescale if the angle between the spin axis of the neutron star companion and the orbital angular momentum is sufficiently large. Nearly two decades later, a new search campaign with the sensitive FAST radio telescope \citep{Wang+vanLeeuwn2025} also yielded no periodic pulsar signals. Ongoing searches for pulsed emission from the binary companion with MeerKAT will be reported elsewhere. 

In the introduction, we mentioned the possibility that PSR~J1906+0746 might instead be of a very similar nature to PSR~J1141$-$6545 \citep{Venkatraman+2020}, with its companion being a massive WD. Given its evolution \citep[described by][]{Tauris+Sennels2000}, such a WD is expected to accrete matter from the pulsar's progenitor, in the same way as the first-formed NS in a DNS system. This potentially spins it up to periods of a few minutes or faster. In the case of PSR~J1141$-$6545, the angular momentum of the WD is large enough to induce detectable spin-orbit coupling in the timing of that pulsar \citep{Venkatraman+2020}. The misalignment between the spin of the WD and the orbital angular momentum causes a precession of both about the total angular momentum of the system. The change of the orientation of the orbital plane and thus of the orbital inclination of the system is detectable as a secular change in the projected semi-major axis, $x = a/c \sin i$ over time, i.e., a measurable $\dot{x}_{\rm obs}$. This spin-orbit coupling is caused, in comparable parts, to the Newtonian spin-orbit coupling (due to the WD's large quadrupole moment) and to the relativistic spin-orbit coupling (the Lense-Thirring effect, which is proportional to the WD's angular momentum).

For PSR~J1906+0746, the 3$\sigma$ measurement of $\dot{x}_{\rm obs}$, $-1.8(6) \times 10^{-13} \, \rm s \, s^{-1}$ (from the DDGR+$\dot{x}$ model in Table~\ref{tab:timing_fitresults}) is similar to the value measured for PSR~J1141$-$6545, \citep{Venkatraman+2020}, raising the possibility that, as in the case of PSR~J1141$-$6545 and PSR~B2303+46 \citep{Church+2006}, its companion is indeed a massive WD formed before the pulsar, though a recycled NS spinning faster than 10\,ms remains a viable alternative.

\begin{figure}
\centering
	\includegraphics[width=\columnwidth]{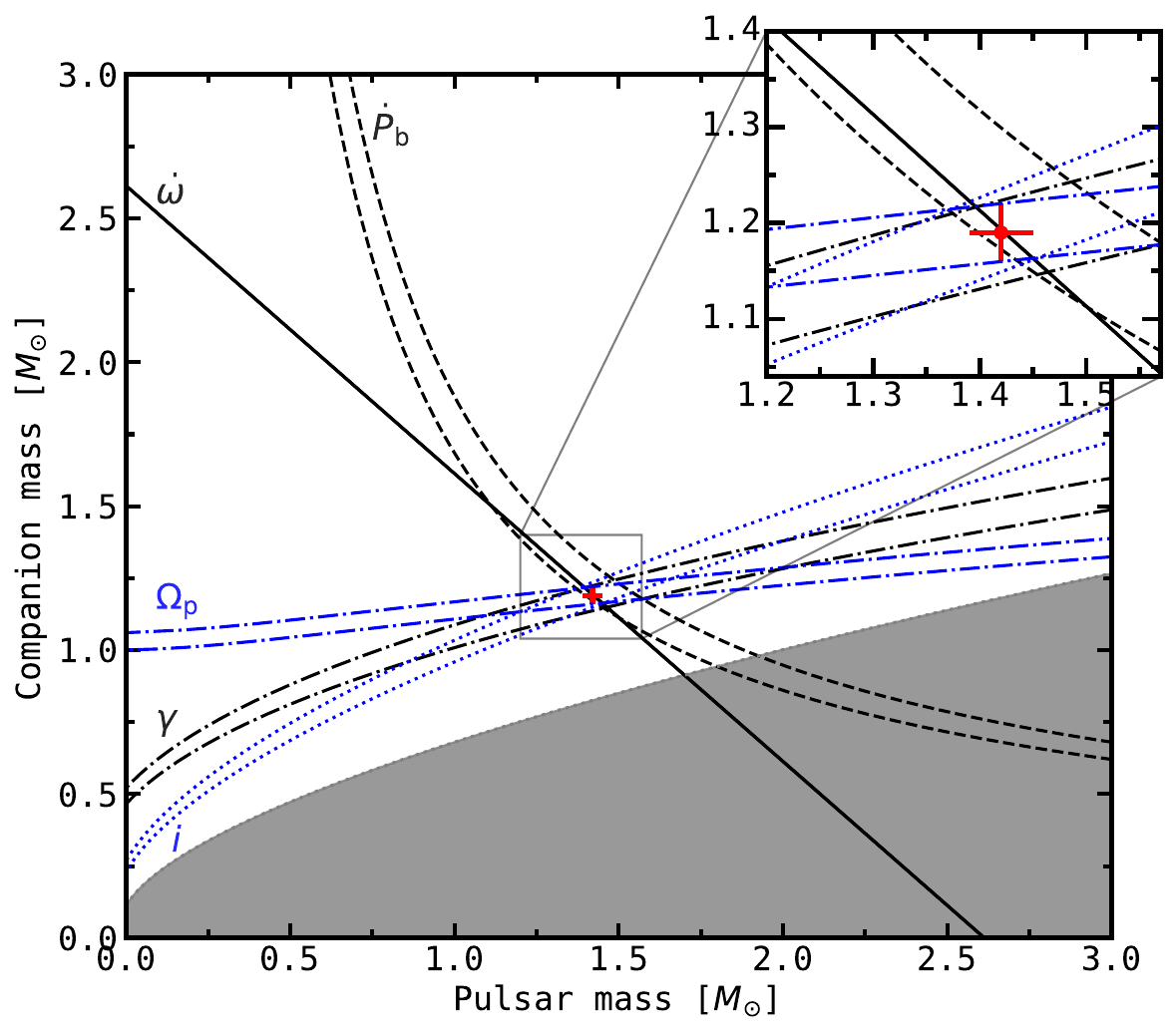}
    \caption{Mass-mass diagram of PSR~J1906+0746 using values from the DDGR+$\dot{x}$ model. The solid lines represent the values of $\dot{\omega}$, the dashed lines show the values of $\dot{P}_{\rm b}$ after including all the contributions (see text for further explanation) and the dash-dotted lines display the values of $\gamma$ as measured by the DD+$\dot{x}$ binary model. The red dot indicates the best-fit value for $M_{\rm p}$ and $M_{\rm c}$ obtained from the DDGR model after fitting for $\dot{x}$. Mass values in the grey region are excluded by the mass function and the condition $i \leq 90^{\circ}$. The dotted and dashed-dotted blue lines display the measured values from $\Omega_{\rm p}$ and $i$  by \citet{Desvignes+2019}.}
    \label{fig:m-m_diagram_xdot}
\end{figure}

Given the correlation of $\dot{x}$ with $\gamma$ \citep{Ridolfi+2019}, the inclusion $\dot{x}$ in the fitting shifts the estimate of $\gamma$ and thus of the individual masses by approximately $3.5\sigma$. The regions of the diagram consistent (in GR) with the measured PK parameters are the bands shown in Figure~\ref{fig:m-m_diagram_xdot}. The companion mass in this case, $1.19(5) \, M_{\odot}$, is still consistent with the idea that the companion is a massive WD, and less consistent with it being the first-formed, recycled NS. Importantly, as shown in Figure~\ref{fig:m-m_diagram_xdot}, the bands associated with the newly fitted PK parameters are still 2-$\sigma$ self-consistent: they still meet in the same region of the mass-mass diagram. This means that the possibility of a large $\dot{x}$ does not contradict with the assumption of the validity of GR.

In this case, we are showing the value of $\dot{P}_{\rm b}$ after accounting for all known contributions (see Section~\ref{s:Pbdot}). Due to the large uncertainties in the pulsar’s proper motion, the $\dot{P}_{\rm b}$ value shown in the diagram includes a Shklovskii contribution considering a total proper motion of 5(14)\,mas\,yr$^{-1}$. This was derived by dividing the measured proper motions in $\alpha$ and $\delta$, along with their uncertainties, by a factor of four, assuming that the system resides in its local co-rotating frame. The fact that using only the directly measured $\dot{P}_{\rm b}$ from the DD timing solution (see Figure~\ref{fig:m-m_diagram}) still yields consistent mass estimates suggests that the additional contributions to $\dot{P}_{\rm b}$ are small. If the system indeed consists of two NSs and there is no measurable $\dot{x}$, then at a distance of 7.4\,kpc, the total contribution from kinematic effects must be minimal.

%%%%%%%%%%%%%%%%%%%%%%%%%%%%%%%%%%%%%%%%%%%%%%%%%%
%%%%%%%%%%%%%%%%%%%% Conclusions %%%%%%%%%%%%%%%%%%

\section{Conclusions}
\label{s:Conclusions}

In this manuscript, we present an analysis of data from multiple radio telescopes of the young relativistic binary PSR~J1906+0746 over a span of 18\,years. By assuming GR, we obtained more precise measurements of the masses of the pulsar and its companion, yielding values of 1.316(5)\,\msun~ for the pulsar and 1.297(5)\,\msun~ for the companion, resulting in a total system mass of 2.6133(1)\,\msun. 

We have also explored the possibility of the companion being a fast-spinning WD, with a 3-$\sigma$ measurement of $\dot{x}$, $\dot{x}_{\rm obs} = -1.8(6)\times 10^{-3}$\,lt-s\,s$^{-1}$.
However, no firm conclusion can be made due to the low significance of the detection. This value of $\dot{x}$ is of a magnitude similar to that observed in PSR~J1141$-$6545. If confirmed, this would, as in the latter case, be due to the coupling of the orbit to the large spin expected for the WD companion, with the observed value being consistent with a spin period of 4\,minutes. Continued timing of the pulsar will be fundamental to confirm this.
If the companion is indeed a fast-spinning WD, PSR~J1906+0746 would offer an excellent laboratory for testing alternative theories of gravity, much like PSRJ1141$-$6545. The $\dot{x}$ could also be explained if the companion were a NS with a spin period smaller than 10\,ms. 

In our analysis, we have detected a large glitch, characterised by a size of $\Delta \nu \sim 41\,\mu$s and a fractional increase of $\Delta \nu/\nu = 5.899(2) \times 10^{-6}$. The glitch exhibited an exponential recovery with a duration of around 100\,days and a degree of recovery with $Q = 0.005$. With a permanent offset in the value of $\dot \nu$ over time following the glitch. 

We have modelled the timing noise as a power-law Gaussian process. Our results suggest a quasi-periodic behaviour with a periodicity of approximately $\sim 2$\,yr. A similar quasi-periodic pattern is also observed in the variation of $\dot{\nu}$, warranting further investigation. Moreover, the PSD suggests a broken power-law with a steep red component transitioning to a flatter red component. Trying to fit for more complicated models, we find that the data are not consistent with stationary power-law noise, which is likely to be a consequence of the complicated nature of its emission. Improved modelling of the timing noise in this system would help to confirm or refute the possibility of a periodicity identified in this work.

Since the discovery of the first binary pulsar, PSR~B1913+16 \cite{Hulse+Taylor1975}, binary pulsars have been instrumental in enabling precise tests of theories of gravity. In particular, tests of the radiative properties of gravity, via the orbital decay of that system \citep{Taylor+Weisberg1982, Taylor+Weisberg1989}, cannot be carried out in the Solar System; they provided the first experimental evidence for the existence of gravitational waves. Since then, the rates of orbital decay observed in binary pulsars have been found to conform to the predictions of GR (\citealp{Kramer+2021}; for a review see \citealp{Freire+Wex2024}).

However, as discussed in Section~\ref{s:Timing}, this requires accurate corrections for the kinematic contributions to the observed variation of the orbital period. For this reason, a precise proper-motion measurement of PSR~B1913+16 would significantly improve the accuracy of these kinematic corrections and, consequently, of the intrinsic orbital-decay measurement, thereby improving the precision of the GR test with this system. More generally, the same considerations apply to other relativistic binaries used for tests of gravity, where accurate astrometric measurements are essential to separate intrinsic orbital decay from kinematic contributions.

These tests are especially powerful for pulsar-WD systems. In such binaries, many alternative theories of gravity predict the emission of dipolar gravitational waves (DGWs) \citep{Eardley1975, Damour+Esposito1996}. Here ``dipolar'' denotes a leading-order dipole radiation channel predicted by a broad class of non-GR theories (e.g., scalar--tensor gravity) in which the two bodies possess different effective gravitational charges (often described in terms of sensitivities or scalar charges). Such dipolar radiation would give rise to an additional contribution to the intrinsic orbital period derivative, $\dot{P}_b^{\rm int}$, beyond that expected from quadrupolar gravitational-wave emission in GR. Therefore, once kinematic contributions have been removed, a comparison between the measured $\dot{P}_b^{\rm int}$ and the GR prediction can be used to place constraints on the presence of DGWs and, more generally, on alternative theories of gravity. Detecting this phenomenon would fundamentally change our understanding of gravitational radiation and would falsify GR. To date, no DGW emission has been detected in any pulsar-WD system (\citealp{Bhat+2008, Freire+2012, Guo+2021}; for a review see \citealp{Freire+Wex2024}), indicating that gravitational waves are, to leading order and to a high degree of purity, quadrupolar as predicted by GR.

A precise proper motion measurement, capable of tightly constraining $\dot{P}_{\rm b,obs}$ without relying on assumptions about kinematic contributions (as illustrated in Figure~\ref{fig:m-m_diagram}), would likely make PSR~J1906+0746 one of the best existing tests of dipolar gravitational radiation in pulsar-WD binaries. Such a measurement has not been possible due to the current very low flux density of the pulsar which is not sufficient for high-precision Very Long Baseline Interferometry (VLBI) astrometry. In principle, this will become feasible with the next-generation telescope Square Kilometre Array (SKA), whose sensitivity will allow the detection of the pulsar at the required S/N for VLBI. However, the detectable emission of the pulsar is expected to disappear by 2028, and will not reappear until between 2070 and 2090, meaning that such a measurement will only become possible once the pulsar is bright enough again and can be observed with SKA-enabled VLBI. This timescale extends well beyond the next few decades. By the time the pulsar re-emerges, the scientific landscape and observational capabilities will likely have advanced considerably, giving place for new possibilities for testing gravity with such systems.

\section*{Acknowledgements}
%People
We thank Chenchen Miao and Pei Wang for assistance with data transport. We also thank Norbert Wex and C.-H. Rosie Chen for insightful discussions and valuable feedback, the referee for a careful reading of the manuscript and constructive comments that improved this work, and the scientific editor, Timothy J. Pearson, for helpful suggestions.
%FAST
This work has used the data from the Five-hundred-meter Aperture Spherical radio Telescope (FAST).  FAST is a Chinese national mega-science facility, operated by the National Astronomical Observatories of Chinese Academy of Sciences (NAOC).
% Arecibo
The Arecibo Observatory was operated by SRI International 
and by University of
Central Florida under a cooperative agreement with the US National Science Foundation, in alliance with Ana G. Mendez-Universidad Metropolitana (UMET), Yang Enterprises, and the Universities Space Research Association (USRA). 
%MeerKAT
The MeerKAT telescope is operated by the South African Radio Astronomy Observatory (SARAO), which is a facility of the National Research Foundation, an agency of the Department of Science and Innovation. SARAO acknowledges the ongoing advice and calibration of GPS systems by the National Metrology Institute of South Africa (NMISA) and the time space reference systems department of the Paris Observatory. PTUSE was developed with support from the Australian SKA Office and Swinburne University of Technology. MeerTime data are housed on the OzSTAR supercomputer at Swinburne University of Technology. The OzSTAR programme receives funding in part from the Astronomy National Collaborative Research Infrastructure Strategy (NCRIS) allocation provided by the Australian Government. 
LV acknowledges financial support from the Dean’s Doctoral Scholar Award from the University of Manchester and partial support from NSF grant AST-1816492.
%Ben
%MK
Pulsar research at Jodrell Bank is supported by a consolidated grant from the UK Science and Technology Facilities Council (STFC).
%MP
GD, PCCF, MK, and VVK acknowledge continuing valuable support from the Max-Planck Society. MK acknowledges significant support from the Max-Planck Society (MPG) and the MPIfR contribution to the PTUSE hardware.
VVK acknowledges financial support from the European Research Council (ERC) starting grant ``COMPACT'' (Grant agreement number 101078094), under the European Union's Horizon Europe research and innovation programme.
% JVL+YYW
JvL and YYW acknowledge funding from 
Vici research programme `ARGO' with project number 639.043.815,
and from CORTEX (NWA.1160.18.316), under the research programme NWA-ORC,
both financed by the Dutch Research Council (NWO).
%IS
Pulsar research at UBC is supported by an NSERC Discovery Grant and by the Canadian Institute for Advanced Research.

\section*{Data Availability}
The pulsar ephemerides, ToAs, and noise models are available upon request from the corresponding authors. The MeerTime observations
are available from the MeerTime data portal, \url{https://pulsars.org.au/}. The FAST data is publicly available (project PT2021\_0001, PI: JvL).

%%%%%%%%%%%%%%%%%%%%%%%%%%%%%%%%%%%%%%%%%%%%%%%%%%

%%%%%%%%%%%%%%%%%%%% References %%%%%%%%%%%%%%%%%%


\bibliographystyle{mnras}
\bibliography{J1906+0746} % if your bibtex file is called example.bib

@ARTICLE{Cordes+2006,
       author = {{Cordes}, J.~M. and {Freire}, P.~C.~C. and {Lorimer}, D.~R. and {Camilo}, F. and {Champion}, D.~J. and {Nice}, D.~J. and {Ramachandran}, R. and {Hessels}, J.~W.~T. and {Vlemmings}, W. and {van Leeuwen}, J. and {Ransom}, S.~M. and {Bhat}, N.~D.~R. and {Arzoumanian}, Z. and {McLaughlin}, M.~A. and {Kaspi}, V.~M. and {Kasian}, L. and {Deneva}, J.~S. and {Reid}, B. and {Chatterjee}, S. and {Han}, J.~L. and {Backer}, D.~C. and {Stairs}, I.~H. and {Deshpande}, A.~A. and {Faucher-Gigu{\`e}re}, C. -A.},
        title = "{Arecibo Pulsar Survey Using ALFA. I. Survey Strategy and First Discoveries}",
      journal = {\apj},
         year = 2006,
        month = jan,
       volume = {637},
       number = {1},
        pages = {446-455},
          doi = {10.1086/498335},
archivePrefix = {arXiv},
       eprint = {astro-ph/0509732},
 primaryClass = {astro-ph},
       adsurl = {https://ui.adsabs.harvard.edu/abs/2006ApJ...637..446C}
}

@ARTICLE{Lorimer+2006,
       author = {{Lorimer}, D.~R. and {Stairs}, I.~H. and {Freire}, P.~C. and {Cordes}, J.~M. and {Camilo}, F. and {Faulkner}, A.~J. and {Lyne}, A.~G. and {Nice}, D.~J. and {Ransom}, S.~M. and {Arzoumanian}, Z. and {Manchester}, R.~N. and {Champion}, D.~J. and {van Leeuwen}, J. and {Mclaughlin}, M.~A. and {Ramachandran}, R. and {Hessels}, J.~W. and {Vlemmings}, W. and {Deshpande}, A.~A. and {Bhat}, N.~D. and {Chatterjee}, S. and {Han}, J.~L. and {Gaensler}, B.~M. and {Kasian}, L. and {Deneva}, J.~S. and {Reid}, B. and {Lazio}, T.~J. and {Kaspi}, V.~M. and {Crawford}, F. and {Lommen}, A.~N. and {Backer}, D.~C. and {Kramer}, M. and {Stappers}, B.~W. and {Hobbs}, G.~B. and {Possenti}, A. and {D'Amico}, N. and {Burgay}, M.},
        title = "{Arecibo Pulsar Survey Using ALFA. II. The Young, Highly Relativistic Binary Pulsar J1906+0746}",
      journal = {\apj},
         year = 2006,
        month = mar,
       volume = {640},
       number = {1},
        pages = {428-434},
          doi = {10.1086/499918},
archivePrefix = {arXiv},
       eprint = {astro-ph/0511523},
 primaryClass = {astro-ph},
       adsurl = {https://ui.adsabs.harvard.edu/abs/2006ApJ...640..428L}
}

@ARTICLE{vanLeeuwen+2006,
       author = {{van Leeuwen}, J. and {Cordes}, J.~M. and {Lorimer}, D.~R. and {Freire}, P.~C.~C. and {Camilo}, F. and {Stairs}, I.~H. and {Nice}, D.~J. and {Champion}, D.~J. and {Ramachandran}, R. and {Faulkner}, A.~J. and {Lyne}, A.~G. and {Ransom}, S.~M. and {Arzoumanian}, Z. and {Manchester}, R.~N. and {McLaughlin}, M.~A. and {Hessels}, J.~W.~T. and {Vlemmings}, W. and {Deshpande}, A.~A. and {Bhat}, N.~D.~R. and {Chatterjee}, S. and {Han}, J.~L. and {Gaensler}, B.~M. and {Kasian}, L. and {Deneva}, J.~S. and {Reid}, B. and {Lazio}, T.~J.~W. and {Kaspi}, V.~M. and {Crawford}, F. and {Lommen}, A.~N. and {Backer}, D.~C. and {Kramer}, M. and {Stappers}, B.~W. and {Hobbs}, G.~B. and {Possenti}, A. and {D'Amico}, N. and {Faucher-Gigu{\`e}re}, C.~A. and {Burgay}, M.},
        title = "{Arecibo and the ALFA Pulsar Survey}",
      journal = {Chin. J. Astron. Astrophys.},
         year = 2006,
        month = dec,
       volume = {6},
       number = {S2},
        pages = {311-318},
          doi = {10.1088/1009-9271/6/S2/58},
          url = {https://dx.doi.org/10.1088/1009-9271/6/S2/58},
       eprint = {astro-ph/0604392},
 primaryClass = {astro-ph},
       adsurl = {https://ui.adsabs.harvard.edu/abs/2006ChJAS...6b.311V}
}

@ARTICLE{vanLeeuwen+2015,
       author = {{van Leeuwen}, J. and {Kasian}, L. and {Stairs}, I.~H. and {Lorimer}, D.~R. and {Camilo}, F. and {Chatterjee}, S. and {Cognard}, I. and {Desvignes}, G. and {Freire}, P.~C.~C. and {Janssen}, G.~H. and {Kramer}, M. and {Lyne}, A.~G. and {Nice}, D.~J. and {Ransom}, S.~M. and {Stappers}, B.~W. and {Weisberg}, J.~M.},
        title = "{The Binary Companion of Young, Relativistic Pulsar J1906+0746}",
      journal = {\apj},
         year = 2015,
        month = jan,
       volume = {798},
       number = {2},
          eid = {118},
        pages = {118},
          doi = {10.1088/0004-637X/798/2/118},
archivePrefix = {arXiv},
       eprint = {1411.1518},
 primaryClass = {astro-ph.SR},
       adsurl = {https://ui.adsabs.harvard.edu/abs/2015ApJ...798..118V}
}

@INPROCEEDINGS{Desvignes+2008,
       author = {{Desvignes}, G. and {Cognard}, I. and {Kramer}, M. and {Lyne}, A. and {Stappers}, B. and {Theureau}, G.},
        title = "{Change in the pulse component separation for PSR J1906+0746}",
    booktitle = {40 Years of Pulsars: Millisecond Pulsars, Magnetars and More},
         year = 2008,
       editor = {{Bassa}, C. and {Wang}, Z. and {Cumming}, A. and {Kaspi}, V.~M.},
       series = {American Institute of Physics Conference Series},
       volume = {983},
        month = feb,
        pages = {482-484},
          doi = {10.1063/1.2900280},
       adsurl = {https://ui.adsabs.harvard.edu/abs/2008AIPC..983..482D}
}

@ARTICLE{Desvignes+2019,
       author = {{Desvignes}, Gregory and {Kramer}, Michael and {Lee}, Kejia and {van Leeuwen}, Joeri and {Stairs}, Ingrid and {Jessner}, Axel and {Cognard}, Isma{\"e}l and {Kasian}, Laura and {Lyne}, Andrew and {Stappers}, Ben W.},
        title = "{Radio emission from a pulsar{\textquoteright}s magnetic pole revealed by general relativity}",
      journal = {Science},
         year = 2019,
        month = sep,
       volume = {365},
       number = {6457},
        pages = {1013-1017},
          doi = {10.1126/science.aav7272},
archivePrefix = {arXiv},
       eprint = {1909.06212},
 primaryClass = {astro-ph.HE},
       adsurl = {https://ui.adsabs.harvard.edu/abs/2019Sci...365.1013D}
}

@ARTICLE{Weltevrede+Johnston2008,
       author = {{Weltevrede}, Patrick and {Johnston}, Simon},
        title = "{Profile and polarization characteristics of energetic pulsars}",
      journal = {\mnras},
         year = 2008,
        month = dec,
       volume = {391},
       number = {3},
        pages = {1210-1226},
          doi = {10.1111/j.1365-2966.2008.13950.x},
archivePrefix = {arXiv},
       eprint = {0809.2438},
 primaryClass = {astro-ph},
       adsurl = {https://ui.adsabs.harvard.edu/abs/2008MNRAS.391.1210W}
}

@ARTICLE{Bailes+2020,
       author = {{Bailes}, M. and {Jameson}, A. and {Abbate}, F. and {Barr}, E.~D. and {Bhat}, N.~D.~R. and {Bondonneau}, L. and {Burgay}, M. and {Buchner}, S.~J. and {Camilo}, F. and {Champion}, D.~J. and {Cognard}, I. and {Demorest}, P.~B. and {Freire}, P.~C.~C. and {Gautam}, T. and {Geyer}, M. and {Griessmeier}, J. -M. and {Guillemot}, L. and {Hu}, H. and {Jankowski}, F. and {Johnston}, S. and {Karastergiou}, A. and {Karuppusamy}, R. and {Kaur}, D. and {Keith}, M.~J. and {Kramer}, M. and {van Leeuwen}, J. and {Lower}, M.~E. and {Maan}, Y. and {McLaughlin}, M.~A. and {Meyers}, B.~W. and {Os{\l}owski}, S. and {Oswald}, L.~S. and {Parthasarathy}, A. and {Pennucci}, T. and {Posselt}, B. and {Possenti}, A. and {Ransom}, S.~M. and {Reardon}, D.~J. and {Ridolfi}, A. and {Schollar}, C.~T.~G. and {Serylak}, M. and {Shaifullah}, G. and {Shamohammadi}, M. and {Shannon}, R.~M. and {Sobey}, C. and {Song}, X. and {Spiewak}, R. and {Stairs}, I.~H. and {Stappers}, B.~W. and {van Straten}, W. and {Szary}, A. and {Theureau}, G. and {Venkatraman Krishnan}, V. and {Weltevrede}, P. and {Wex}, N. and {Abbott}, T.~D. and {Adams}, G.~B. and {Burger}, J.~P. and {Gamatham}, R.~R.~G. and {Gouws}, M. and {Horn}, D.~M. and {Hugo}, B. and {Joubert}, A.~F. and {Manley}, J.~R. and {McAlpine}, K. and {Passmoor}, S.~S. and {Peens-Hough}, A. and {Ramudzuli}, Z.~R. and {Rust}, A. and {Salie}, S. and {Schwardt}, L.~C. and {Siebrits}, R. and {Van Tonder}, G. and {Van Tonder}, V. and {Welz}, M.~G.},
        title = "{The MeerKAT telescope as a pulsar facility: System verification and early science results from MeerTime}",
      journal = {\pasa},
         year = 2020,
        month = jul,
       volume = {37},
          eid = {e028},
        pages = {e028},
          doi = {10.1017/pasa.2020.19},
archivePrefix = {arXiv},
       eprint = {2005.14366},
 primaryClass = {astro-ph.IM},
       adsurl = {https://ui.adsabs.harvard.edu/abs/2020PASA...37...28B}
}

@ARTICLE{Hotan+2004,
       author = {{Hotan}, A.~W. and {van Straten}, W. and {Manchester}, R.~N.},
        title = "{PSRCHIVE and PSRFITS: An Open Approach to Radio Pulsar Data Storage and Analysis}",
      journal = {\pasa},
         year = 2004,
        month = jan,
       volume = {21},
       number = {3},
        pages = {302-309},
          doi = {10.1071/AS04022},
archivePrefix = {arXiv},
       eprint = {astro-ph/0404549},
 primaryClass = {astro-ph},
       adsurl = {https://ui.adsabs.harvard.edu/abs/2004PASA...21..302H}
}

@MISC{Lazarus+2020,
       author = {{Lazarus}, P. and {Karuppusamy}, R. and {Graikou}, E. and {Caballero}, R.~N. and {Champion}, D.~J. and {Lee}, K.~J. and {Verbiest}, J.~P.~W. and {Kramer}, M.},
        title = "{CoastGuard: Automated timing data reduction pipeline}",
 howpublished = {Astrophysics Source Code Library, record ascl:2003.008},
         year = 2020,
        month = mar,
          eid = {ascl:2003.008},
        pages = {ascl:2003.008},
archivePrefix = {ascl},
       eprint = {2003.008},
       adsurl = {https://ui.adsabs.harvard.edu/abs/2020ascl.soft03008L}
}

@ARTICLE{Serylak+2021,
       author = {{Serylak}, M. and {Johnston}, S. and {Kramer}, M. and {Buchner}, S. and {Karastergiou}, A. and {Keith}, M.~J. and {Parthasarathy}, A. and {Weltevrede}, P. and {Bailes}, M. and {Barr}, E.~D. and {Camilo}, F. and {Geyer}, M. and {Hugo}, B.~V. and {Jameson}, A. and {Reardon}, D.~J. and {Shannon}, R.~M. and {Spiewak}, R. and {van Straten}, W. and {Venkatraman Krishnan}, V.},
        title = "{The thousand-pulsar-array programme on MeerKAT IV: Polarization properties of young, energetic pulsars}",
      journal = {\mnras},
         year = 2021,
        month = aug,
       volume = {505},
       number = {3},
        pages = {4483-4495},
          doi = {10.1093/mnras/staa2811},
archivePrefix = {arXiv},
       eprint = {2009.05797},
 primaryClass = {astro-ph.HE},
       adsurl = {https://ui.adsabs.harvard.edu/abs/2021MNRAS.505.4483S}
}

@ARTICLE{Park+2021,
       author = {{Park}, Ryan S. and {Folkner}, William M. and {Williams}, James G. and {Boggs}, Dale H.},
        title = "{The JPL Planetary and Lunar Ephemerides DE440 and DE441}",
      journal = {\aj},
         year = 2021,
        month = mar,
       volume = {161},
       number = {3},
          eid = {105},
        pages = {105},
          doi = {10.3847/1538-3881/abd414},
       adsurl = {https://ui.adsabs.harvard.edu/abs/2021AJ....161..105P}
}

@ARTICLE{Kramer+2021,
       author = {{Kramer}, M. and {Stairs}, I.~H. and {Venkatraman Krishnan}, V. and {Freire}, P.~C.~C. and {Abbate}, F. and {Bailes}, M. and {Burgay}, M. and {Buchner}, S. and {Champion}, D.~J. and {Cognard}, I. and {Gautam}, T. and {Geyer}, M. and {Guillemot}, L. and {Hu}, H. and {Janssen}, G. and {Lower}, M.~E. and {Parthasarathy}, A. and {Possenti}, A. and {Ransom}, S. and {Reardon}, D.~J. and {Ridolfi}, A. and {Serylak}, M. and {Shannon}, R.~M. and {Spiewak}, R. and {Theureau}, G. and {van Straten}, W. and {Wex}, N. and {Oswald}, L.~S. and {Posselt}, B. and {Sobey}, C. and {Barr}, E.~D. and {Camilo}, F. and {Hugo}, B. and {Jameson}, A. and {Johnston}, S. and {Karastergiou}, A. and {Keith}, M. and {Os{\l}owski}, S.},
        title = "{The relativistic binary programme on MeerKAT: science objectives and first results}",
      journal = {\mnras},
         year = 2021,
        month = jun,
       volume = {504},
       number = {2},
        pages = {2094-2114},
          doi = {10.1093/mnras/stab375},
archivePrefix = {arXiv},
       eprint = {2102.05160},
 primaryClass = {astro-ph.HE},
       adsurl = {https://ui.adsabs.harvard.edu/abs/2021MNRAS.504.2094K}
}

@ARTICLE{Hobbs+2006,
       author = {{Hobbs}, G. and {Edwards}, R. and {Manchester}, R.},
        title = "{TEMPO2: a New Pulsar Timing Package}",
      journal = {Chinese Journal of Astronomy and Astrophysics Supplement},
         year = 2006,
        month = dec,
       volume = {6},
       number = {S2},
        pages = {189-192},
       adsurl = {https://ui.adsabs.harvard.edu/abs/2006ChJAS...6b.189H}
}

@ARTICLE{Barker+OConnell1975,
       author = {{Barker}, B.~M. and {O'Connell}, R.~F.},
        title = "{Relativistic effects in the binary pulsar PSR 1913+16.}",
      journal = {\apjl},
         year = 1975,
        month = jul,
       volume = {199},
        pages = {L25},
          doi = {10.1086/181840},
       adsurl = {https://ui.adsabs.harvard.edu/abs/1975ApJ...199L..25B}
}

@ARTICLE{Shemar+Lyne1996,
       author = {{Shemar}, S.~L. and {Lyne}, A.~G.},
        title = "{Observations of pulsar glitches}",
      journal = {\mnras},
         year = 1996,
        month = sep,
       volume = {282},
       number = {2},
        pages = {677-690},
          doi = {10.1093/mnras/282.2.677},
       adsurl = {https://ui.adsabs.harvard.edu/abs/1996MNRAS.282..677S}
}

@ARTICLE{Hobbs+2004,
       author = {{Hobbs}, G. and {Lyne}, A.~G. and {Kramer}, M. and {Martin}, C.~E. and {Jordan}, C.},
        title = "{Long-term timing observations of 374 pulsars}",
      journal = {\mnras},
         year = 2004,
        month = oct,
       volume = {353},
       number = {4},
        pages = {1311-1344},
          doi = {10.1111/j.1365-2966.2004.08157.x},
       adsurl = {https://ui.adsabs.harvard.edu/abs/2004MNRAS.353.1311H}
}

@ARTICLE{Manchester+2013,
       author = {{Manchester}, R.~N. and {Hobbs}, G. and {Bailes}, M. and {Coles}, W.~A. and {van Straten}, W. and {Keith}, M.~J. and {Shannon}, R.~M. and {Bhat}, N.~D.~R. and {Brown}, A. and {Burke-Spolaor}, S.~G. and {Champion}, D.~J. and {Chaudhary}, A. and {Edwards}, R.~T. and {Hampson}, G. and {Hotan}, A.~W. and {Jameson}, A. and {Jenet}, F.~A. and {Kesteven}, M.~J. and {Khoo}, J. and {Kocz}, J. and {Maciesiak}, K. and {Oslowski}, S. and {Ravi}, V. and {Reynolds}, J.~R. and {Sarkissian}, J.~M. and {Verbiest}, J.~P.~W. and {Wen}, Z.~L. and {Wilson}, W.~E. and {Yardley}, D. and {Yan}, W.~M. and {You}, X.~P.},
        title = "{The Parkes Pulsar Timing Array Project}",
      journal = {\pasa},
         year = 2013,
        month = jan,
       volume = {30},
          eid = {e017},
        pages = {e017},
          doi = {10.1017/pasa.2012.017},
archivePrefix = {arXiv},
       eprint = {1210.6130},
 primaryClass = {astro-ph.IM},
       adsurl = {https://ui.adsabs.harvard.edu/abs/2013PASA...30...17M}
}

@ARTICLE{Keith+Nitu2023,
       author = {{Keith}, Michael J. and {Ni{\c{t}}u}, Iuliana C.},
        title = "{Impact of quasi-periodic and steep-spectrum timing noise on the measurement of pulsar timing parameters}",
      journal = {\mnras},
         year = 2023,
        month = aug,
       volume = {523},
       number = {3},
        pages = {4603-4614},
          doi = {10.1093/mnras/stad1713},
archivePrefix = {arXiv},
       eprint = {2306.03529},
 primaryClass = {astro-ph.HE},
       adsurl = {https://ui.adsabs.harvard.edu/abs/2023MNRAS.523.4603K}
}

@ARTICLE{Lentati+2014,
       author = {{Lentati}, L. and {Alexander}, P. and {Hobson}, M.~P. and {Feroz}, F. and {van Haasteren}, R. and {Lee}, K.~J. and {Shannon}, R.~M.},
        title = "{TEMPONEST: a Bayesian approach to pulsar timing analysis}",
      journal = {\mnras},
         year = 2014,
        month = jan,
       volume = {437},
       number = {3},
        pages = {3004-3023},
          doi = {10.1093/mnras/stt2122},
archivePrefix = {arXiv},
       eprint = {1310.2120},
 primaryClass = {astro-ph.IM},
       adsurl = {https://ui.adsabs.harvard.edu/abs/2014MNRAS.437.3004L}
}

@ARTICLE{Foreman-Mackey+2013,
       author = {{Foreman-Mackey}, Daniel and {Hogg}, David W. and {Lang}, Dustin and {Goodman}, Jonathan},
        title = "{emcee: The MCMC Hammer}",
      journal = {\pasp},
         year = 2013,
        month = mar,
       volume = {125},
       number = {925},
        pages = {306},
          doi = {10.1086/670067},
archivePrefix = {arXiv},
       eprint = {1202.3665},
 primaryClass = {astro-ph.IM},
       adsurl = {https://ui.adsabs.harvard.edu/abs/2013PASP..125..306F}
}

@ARTICLE{Yu+2013,
       author = {{Yu}, M. and {Manchester}, R.~N. and {Hobbs}, G. and {Johnston}, S. and {Kaspi}, V.~M. and {Keith}, M. and {Lyne}, A.~G. and {Qiao}, G.~J. and {Ravi}, V. and {Sarkissian}, J.~M. and {Shannon}, R. and {Xu}, R.~X.},
        title = "{Detection of 107 glitches in 36 southern pulsars}",
      journal = {\mnras},
         year = 2013,
        month = feb,
       volume = {429},
       number = {1},
        pages = {688-724},
          doi = {10.1093/mnras/sts366},
archivePrefix = {arXiv},
       eprint = {1211.2035},
 primaryClass = {astro-ph.HE},
       adsurl = {https://ui.adsabs.harvard.edu/abs/2013MNRAS.429..688Y}
}

@MISC{Ellis+2019,
       author = {{Ellis}, Justin A. and {Vallisneri}, Michele and {Taylor}, Stephen R. and {Baker}, Paul T.},
        title = "{ENTERPRISE: Enhanced Numerical Toolbox Enabling a Robust PulsaR Inference SuitE}",
 howpublished = {Astrophysics Source Code Library, record ascl:1912.015},
         year = 2019,
        month = dec,
          eid = {ascl:1912.015},
        pages = {ascl:1912.015},
archivePrefix = {ascl},
       eprint = {1912.015},
       adsurl = {https://ui.adsabs.harvard.edu/abs/2019ascl.soft12015E}
}

@MISC{Keith+2022,
       author = {{Keith}, Michael John and {Ni{\c{t}}u}, Iuliana C. and {Liu}, Yang},
        title = "{run\_enterprise}",
 howpublished = {Zenodo},
         year = 2022,
        month = feb,
          eid = {10.5281/zenodo.6046212},
          doi = {10.5281/zenodo.6046212},
      version = {2022.02.1},
    publisher = {Zenodo},
       adsurl = {https://ui.adsabs.harvard.edu/abs/2022zndo...6046212K}
}

@ARTICLE{Shklovskii1970,
       author = {{Shklovskii}, I.~S.},
        title = "{Possible Causes of the Secular Increase in Pulsar Periods.}",
      journal = {\sovast},
         year = 1970,
        month = feb,
       volume = {13},
        pages = {562},
       adsurl = {https://ui.adsabs.harvard.edu/abs/1970SvA....13..562S}
}

@ARTICLE{Lazaridis+2009,
       author = {{Lazaridis}, K. and {Wex}, N. and {Jessner}, A. and {Kramer}, M. and {Stappers}, B.~W. and {Janssen}, G.~H. and {Desvignes}, G. and {Purver}, M.~B. and {Cognard}, I. and {Theureau}, G. and {Lyne}, A.~G. and {Jordan}, C.~A. and {Zensus}, J.~A.},
        title = "{Generic tests of the existence of the gravitational dipole radiation and the variation of the gravitational constant}",
      journal = {\mnras},
         year = 2009,
        month = dec,
       volume = {400},
       number = {2},
        pages = {805-814},
          doi = {10.1111/j.1365-2966.2009.15481.x},
archivePrefix = {arXiv},
       eprint = {0908.0285},
 primaryClass = {astro-ph.GA},
       adsurl = {https://ui.adsabs.harvard.edu/abs/2009MNRAS.400..805L}
}

@ARTICLE{Grav+2021,
       author = {{GRAVITY Collaboration} and {Abuter}, R. and {Amorim}, A. and {Baub{\"o}ck}, M. and {Berger}, J.~P. and {Bonnet}, H. and {Brandner}, W. and {Cl{\'e}net}, Y. and {Davies}, R. and {de Zeeuw}, P.~T. and {Dexter}, J. and {Dallilar}, Y. and {Drescher}, A. and {Eckart}, A. and {Eisenhauer}, F. and {F{\"o}rster Schreiber}, N.~M. and {Garcia}, P. and {Gao}, F. and {Gendron}, E. and {Genzel}, R. and {Gillessen}, S. and {Habibi}, M. and {Haubois}, X. and {Hei{\ss}el}, G. and {Henning}, T. and {Hippler}, S. and {Horrobin}, M. and {Jim{\'e}nez-Rosales}, A. and {Jochum}, L. and {Jocou}, L. and {Kaufer}, A. and {Kervella}, P. and {Lacour}, S. and {Lapeyr{\`e}re}, V. and {Le Bouquin}, J. -B. and {L{\'e}na}, P. and {Lutz}, D. and {Nowak}, M. and {Ott}, T. and {Paumard}, T. and {Perraut}, K. and {Perrin}, G. and {Pfuhl}, O. and {Rabien}, S. and {Rodr{\'\i}guez-Coira}, G. and {Shangguan}, J. and {Shimizu}, T. and {Scheithauer}, S. and {Stadler}, J. and {Straub}, O. and {Straubmeier}, C. and {Sturm}, E. and {Tacconi}, L.~J. and {Vincent}, F. and {von Fellenberg}, S. and {Waisberg}, I. and {Widmann}, F. and {Wieprecht}, E. and {Wiezorrek}, E. and {Woillez}, J. and {Yazici}, S. and {Young}, A. and {Zins}, G.},
        title = "{Improved GRAVITY astrometric accuracy from modeling optical aberrations}",
      journal = {\aap},
         year = 2021,
        month = mar,
       volume = {647},
          eid = {A59},
        pages = {A59},
          doi = {10.1051/0004-6361/202040208},
archivePrefix = {arXiv},
       eprint = {2101.12098},
 primaryClass = {astro-ph.GA},
       adsurl = {https://ui.adsabs.harvard.edu/abs/2021A&A...647A..59G}
}

@ARTICLE{Guo+2021,
       author = {{Guo}, Y.~J. and {Freire}, P.~C.~C. and {Guillemot}, L. and {Kramer}, M. and {Zhu}, W.~W. and {Wex}, N. and {McKee}, J.~W. and {Deller}, A. and {Ding}, H. and {Kaplan}, D.~L. and {Stappers}, B. and {Cognard}, I. and {Miao}, X. and {Haase}, L. and {Keith}, M. and {Ransom}, S.~M. and {Theureau}, G.},
        title = "{PSR J2222{\ensuremath{-}}0137. I. Improved physical parameters for the system}",
      journal = {\aap},
         year = 2021,
        month = oct,
       volume = {654},
          eid = {A16},
        pages = {A16},
          doi = {10.1051/0004-6361/202141450},
archivePrefix = {arXiv},
       eprint = {2107.09474},
 primaryClass = {astro-ph.HE},
       adsurl = {https://ui.adsabs.harvard.edu/abs/2021A&A...654A..16G}
}

@ARTICLE{Damour+Deruelle1985,
       author = {{Damour}, T. and {Deruelle}, N.},
        title = "{General relativistic celestial mechanics of binary systems. I. The post-Newtonian motion.}",
      journal = {Annales de L'Institut Henri Poincare Section (A) Physique Theorique},
         year = 1985,
        month = jan,
       volume = {43},
       number = {1},
        pages = {107-132},
       adsurl = {https://ui.adsabs.harvard.edu/abs/1985AIHPA..43..107D}
}

@ARTICLE{Damour+Deruelle1986,
       author = {{Damour}, T. and {Deruelle}, N.},
        title = "{General relativistic celestial mechanics of binary systems. II. The post-Newtonian timing formula.}",
      journal = {Annales de L'Institut Henri Poincare Section (A) Physique Theorique},
         year = 1986,
        month = jan,
       volume = {44},
       number = {3},
        pages = {263-292},
       adsurl = {https://ui.adsabs.harvard.edu/abs/1986AIHPA..44..263D}
}

@ARTICLE{Taylor+Weisberg1989,
       author = {{Taylor}, J.~H. and {Weisberg}, J.~M.},
        title = "{Further Experimental Tests of Relativistic Gravity Using the Binary Pulsar PSR 1913+16}",
      journal = {\apj},
         year = 1989,
        month = oct,
       volume = {345},
        pages = {434},
          doi = {10.1086/167917},
       adsurl = {https://ui.adsabs.harvard.edu/abs/1989ApJ...345..434T}
}

@ARTICLE{Weltevrede+2010,
       author = {{Weltevrede}, P. and {Johnston}, S. and {Manchester}, R.~N. and {Bhat}, R. and {Burgay}, M. and {Champion}, D. and {Hobbs}, G.~B. and {K{\i}z{\i}ltan}, B. and {Keith}, M. and {Possenti}, A. and {Reynolds}, J.~E. and {Watters}, K.},
        title = "{Pulsar Timing with the Parkes Radio Telescope for the Fermi Mission}",
      journal = {\pasa},
         year = 2010,
        month = mar,
       volume = {27},
       number = {1},
        pages = {64-75},
          doi = {10.1071/AS09054},
archivePrefix = {arXiv},
       eprint = {0909.5510},
 primaryClass = {astro-ph.GA},
       adsurl = {https://ui.adsabs.harvard.edu/abs/2010PASA...27...64W}
}

@ARTICLE{Nitu+2022,
       author = {{Ni{\c{t}}u}, Iuliana C. and {Keith}, Michael J. and {Stappers}, Ben W. and {Lyne}, Andrew G. and {Mickaliger}, Mitchell B.},
        title = "{A search for planetary companions around 800 pulsars from the Jodrell Bank pulsar timing programme}",
      journal = {\mnras},
         year = 2022,
        month = may,
       volume = {512},
       number = {2},
        pages = {2446-2459},
          doi = {10.1093/mnras/stac593},
archivePrefix = {arXiv},
       eprint = {2203.01136},
 primaryClass = {astro-ph.EP},
       adsurl = {https://ui.adsabs.harvard.edu/abs/2022MNRAS.512.2446N}
}

@ARTICLE{Lyne+2010,
       author = {{Lyne}, Andrew and {Hobbs}, George and {Kramer}, Michael and {Stairs}, Ingrid and {Stappers}, Ben},
        title = "{Switched Magnetospheric Regulation of Pulsar Spin-Down}",
      journal = {Science},
         year = 2010,
        month = jul,
       volume = {329},
       number = {5990},
        pages = {408},
          doi = {10.1126/science.1186683},
archivePrefix = {arXiv},
       eprint = {1006.5184},
 primaryClass = {astro-ph.GA},
       adsurl = {https://ui.adsabs.harvard.edu/abs/2010Sci...329..408L}
}

@ARTICLE{vanKerkwijk+Kulkarni1999,
       author = {{van Kerkwijk}, M.~H. and {Kulkarni}, S.~R.},
        title = "{A Massive White Dwarf Companion to the Eccentric Binary Pulsar System PSR B2303+46}",
      journal = {\apjl},
         year = 1999,
        month = may,
       volume = {516},
       number = {1},
        pages = {L25-L28},
          doi = {10.1086/311991},
archivePrefix = {arXiv},
       eprint = {astro-ph/9901149},
 primaryClass = {astro-ph},
       adsurl = {https://ui.adsabs.harvard.edu/abs/1999ApJ...516L..25V}
}

@INPROCEEDINGS{Dowd+2000,
       author = {{Dowd}, A. and {Sisk}, W. and {Hagen}, J.},
        title = "{WAPP --- Wideband Arecibo Pulsar Processor}",
    booktitle = {IAU Colloq. 177: Pulsar Astronomy - 2000 and Beyond},
         year = 2000,
       editor = {{Kramer}, M. and {Wex}, N. and {Wielebinski}, R.},
       series = {Astronomical Society of the Pacific Conference Series},
       volume = {202},
        month = jan,
        pages = {275-276},
       adsurl = {https://ui.adsabs.harvard.edu/abs/2000ASPC..202..275D}
}

@INPROCEEDINGS{Demorest+2004,
       author = {{Demorest}, P. and {Ramachandran}, R. and {Backer}, D. and {Ferdman}, R. and {Stairs}, I. and {Nice}, D.},
        title = "{Precision Pulsar Timing and Gravity Waves: Recent Advances in Instrumentation}",
    booktitle = {American Astronomical Society Meeting Abstracts},
         year = 2004,
       series = {American Astronomical Society Meeting Abstracts},
       volume = {205},
        month = dec,
          eid = {149.01},
        pages = {149.01},
       adsurl = {https://ui.adsabs.harvard.edu/abs/2004AAS...20514901D}
}

@ARTICLE{Kaspi+2000,
       author = {{Kaspi}, V.~M. and {Lyne}, A.~G. and {Manchester}, R.~N. and {Crawford}, F. and {Camilo}, F. and {Bell}, J.~F. and {D'Amico}, N. and {Stairs}, I.~H. and {McKay}, N.~P.~F. and {Morris}, D.~J. and {Possenti}, A.},
        title = "{Discovery of a Young Radio Pulsar in a Relativistic Binary Orbit}",
      journal = {\apj},
         year = 2000,
        month = nov,
       volume = {543},
       number = {1},
        pages = {321-327},
          doi = {10.1086/317103},
archivePrefix = {arXiv},
       eprint = {astro-ph/0005214},
 primaryClass = {astro-ph},
       adsurl = {https://ui.adsabs.harvard.edu/abs/2000ApJ...543..321K}
}

@ARTICLE{Tauris+Sennels2000,
       author = {{Tauris}, T.~M. and {Sennels}, T.},
        title = "{Formation of the binary pulsars PSR B2303+46 and PSR J1141-6545. Young neutron stars with old white dwarf companions}",
      journal = {\aap},
         year = 2000,
        month = mar,
       volume = {355},
        pages = {236-244},
          doi = {10.48550/arXiv.astro-ph/9909149},
archivePrefix = {arXiv},
       eprint = {astro-ph/9909149},
 primaryClass = {astro-ph},
       adsurl = {https://ui.adsabs.harvard.edu/abs/2000A&A...355..236T}
}

@ARTICLE{Wang+2000,
       author = {{Wang}, N. and {Manchester}, R.~N. and {Pace}, R.~T. and {Bailes}, M. and {Kaspi}, V.~M. and {Stappers}, B.~W. and {Lyne}, A.~G.},
        title = "{Glitches in southern pulsars}",
      journal = {\mnras},
         year = 2000,
        month = oct,
       volume = {317},
       number = {4},
        pages = {843-860},
          doi = {10.1046/j.1365-8711.2000.03713.x},
archivePrefix = {arXiv},
       eprint = {astro-ph/0005561},
 primaryClass = {astro-ph},
       adsurl = {https://ui.adsabs.harvard.edu/abs/2000MNRAS.317..843W}
}

@ARTICLE{Damour+Taylor1991,
       author = {{Damour}, Thibault and {Taylor}, J.~H.},
        title = "{On the Orbital Period Change of the Binary Pulsar PSR 1913+16}",
      journal = {\apj},
         year = 1991,
        month = jan,
       volume = {366},
        pages = {501},
          doi = {10.1086/169585},
       adsurl = {https://ui.adsabs.harvard.edu/abs/1991ApJ...366..501D}
}

@ARTICLE{Nice+Taylor1995,
       author = {{Nice}, D.~J. and {Taylor}, J.~H.},
        title = "{PSR J2019+2425 and PSR J2322+2057 and the Proper Motions of Millisecond Pulsars}",
      journal = {\apj},
         year = 1995,
        month = mar,
       volume = {441},
        pages = {429},
          doi = {10.1086/175367},
       adsurl = {https://ui.adsabs.harvard.edu/abs/1995ApJ...441..429N}
}

@ARTICLE{Kramer+Wex2009,
       author = {{Kramer}, M and {Wex}, N},
        title = "{TOPICAL REVIEW:  The double pulsar system: a unique laboratory for gravity}",
      journal = {Classical and Quantum Gravity},
         year = 2009,
        month = apr,
       volume = {26},
       number = {7},
          eid = {073001},
        pages = {073001},
          doi = {10.1088/0264-9381/26/7/073001},
       adsurl = {https://ui.adsabs.harvard.edu/abs/2009CQGra..26g3001K}
}

@ARTICLE{Wang+2025a,
       author = {{Wang}, Y.~Y. and {van Leeuwen}, J. and {Desvignes}, G. and {et al.}},
        title = "{1906 Beam Mapping}",
      journal = {\aap {\it ~in prep.}},
         year = 2026,
         note = {}
}

@ARTICLE{Nan+2011,
       author = {{Nan}, Rendong and {Li}, Di and {Jin}, Chengjin and {Wang}, Qiming and
         {Zhu}, Lichun and {Zhu}, Wenbai and {Zhang}, Haiyan and {Yue}, Youling and
         {Qian}, Lei},
        title = "{The Five-Hundred Aperture Spherical Radio Telescope (fast) Project}",
      journal = {International Journal of Modern Physics D},
         year = 2011,
        month = jan,
       volume = {20},
       number = {6},
        pages = {989-1024},
          doi = {10.1142/S0218271811019335},
archivePrefix = {arXiv},
       eprint = {1105.3794},
 primaryClass = {astro-ph.IM},
       adsurl = {https://ui.adsabs.harvard.edu/abs/2011IJMPD..20..989N}
}

@ARTICLE{Morello+2019,
       author = {{Morello}, V. and {Barr}, E.~D. and {Cooper}, S. and {Bailes}, M. and {Bates}, S. and {Bhat}, N.~D.~R. and {Burgay}, M. and {Burke-Spolaor}, S. and {Cameron}, A.~D. and {Champion}, D.~J. and {Eatough}, R.~P. and {Flynn}, C.~M.~L. and {Jameson}, A. and {Johnston}, S. and {Keith}, M.~J. and {Keane}, E.~F. and {Kramer}, M. and {Levin}, L. and {Ng}, C. and {Petroff}, E. and {Possenti}, A. and {Stappers}, B.~W. and {van Straten}, W. and {Tiburzi}, C.},
        title = "{The High Time Resolution Universe survey - XIV. Discovery of 23 pulsars through GPU-accelerated reprocessing}",
      journal = {\mnras},
         year = 2019,
        month = mar,
       volume = {483},
       number = {3},
        pages = {3673-3685},
          doi = {10.1093/mnras/sty3328},
archivePrefix = {arXiv},
       eprint = {1811.04929},
 primaryClass = {astro-ph.IM},
       adsurl = {https://ui.adsabs.harvard.edu/abs/2019MNRAS.483.3673M}
}

@ARTICLE{Parthasarathy+2019,
       author = {{Parthasarathy}, A. and {Shannon}, R.~M. and {Johnston}, S. and {Lentati}, L. and {Bailes}, M. and {Dai}, S. and {Kerr}, M. and {Manchester}, R.~N. and {Os{\l}owski}, S. and {Sobey}, C. and {van Straten}, W. and {Weltevrede}, P.},
        title = "{Timing of young radio pulsars - I. Timing noise, periodic modulation, and proper motion}",
      journal = {\mnras},
         year = 2019,
        month = nov,
       volume = {489},
       number = {3},
        pages = {3810-3826},
          doi = {10.1093/mnras/stz2383},
archivePrefix = {arXiv},
       eprint = {1908.11709},
 primaryClass = {astro-ph.HE},
       adsurl = {https://ui.adsabs.harvard.edu/abs/2019MNRAS.489.3810P}
}

@BOOK{Lorimer+Kramer2004,
       author = {{Lorimer}, D.~R. and {Kramer}, M.},
        title = "{Handbook of Pulsar Astronomy}",
         year = 2004,
        publisher = {Cambridge University Press},
        series = {Cambridge Observing Handbooks for Research Astronomers, Series},
       volume = {4},
       adsurl = {https://ui.adsabs.harvard.edu/abs/2004hpa..book.....L}
}

@ARTICLE{Kopeikin1996,
       author = {{Kopeikin}, S.~M.},
        title = "{Proper Motion of Binary Pulsars as a Source of Secular Variations of Orbital Parameters}",
      journal = {\apjl},
         year = 1996,
        month = aug,
       volume = {467},
        pages = {L93},
          doi = {10.1086/310201},
       adsurl = {https://ui.adsabs.harvard.edu/abs/1996ApJ...467L..93K}
}

@INPROCEEDINGS{Arzoumanian+1996,
       author = {{Arzoumanian}, Z. and {Joshi}, K. and {Rasio}, F.~A. and {Thorsett}, S.~E.},
        title = "{Orbital Parameters of the PSR B1620-26 Triple System}",
    booktitle = {IAU Colloq. 160: Pulsars: Problems and Progress},
         year = 1996,
       editor = {{Johnston}, S. and {Walker}, M.~A. and {Bailes}, M.},
       series = {Astronomical Society of the Pacific Conference Series},
       volume = {105},
        month = jan,
        pages = {525-530},
          doi = {10.48550/arXiv.astro-ph/9605141},
archivePrefix = {arXiv},
       eprint = {astro-ph/9605141},
 primaryClass = {astro-ph},
       adsurl = {https://ui.adsabs.harvard.edu/abs/1996ASPC..105..525A}
}

@ARTICLE{Damour+Taylor1992,
       author = {{Damour}, Thibault and {Taylor}, J.~H.},
        title = "{Strong-field tests of relativistic gravity and binary pulsars}",
      journal = {\prd},
         year = 1992,
        month = mar,
       volume = {45},
       number = {6},
        pages = {1840-1868},
          doi = {10.1103/PhysRevD.45.1840},
       adsurl = {https://ui.adsabs.harvard.edu/abs/1992PhRvD..45.1840D}
}

@ARTICLE{Boshkayev+2017,
       author = {{Boshkayev}, K. and {Quevedo}, H. and {Zhami}, B.},
        title = "{I -Love- Q relations for white dwarf stars}",
      journal = {\mnras},
         year = 2017,
        month = feb,
       volume = {464},
       number = {4},
        pages = {4349-4359},
          doi = {10.1093/mnras/stw2614},
       adsurl = {https://ui.adsabs.harvard.edu/abs/2017MNRAS.464.4349B}
}

@ARTICLE{Wex+1998,
       author = {{Wex}, N. and {Johnston}, S. and {Manchester}, R.~N. and {Lyne}, A.~G. and {Stappers}, B.~W. and {Bailes}, M.},
        title = "{Timing models for the long orbital period binary pulsar PSR B1259-63}",
      journal = {\mnras},
         year = 1998,
        month = aug,
       volume = {298},
       number = {4},
        pages = {997-1004},
          doi = {10.1046/j.1365-8711.1998.01700.x},
archivePrefix = {arXiv},
       eprint = {astro-ph/9803182},
 primaryClass = {astro-ph},
       adsurl = {https://ui.adsabs.harvard.edu/abs/1998MNRAS.298..997W}
}

@ARTICLE{Ridolfi+2019,
    author = {{Ridolfi}, A. and {Freire}, P.~C.~C. and {Gupta}, Y. and {Ransom}, S.~M.},
    title = "{Upgraded Giant Metrewave Radio Telescope timing of NGC 1851A: a possible millisecond pulsar - neutron star system}",
    journal = {\mnras},
    year = 2019,
    month = dec,
    volume = {490},
    number = {3},
    pages = {3860-3874},
    doi = {10.1093/mnras/stz2645},
archivePrefix = {arXiv},
       eprint = {1909.06163},
 primaryClass = {astro-ph.HE},
       adsurl = {https://ui.adsabs.harvard.edu/abs/2019MNRAS.490.3860R}
}

@ARTICLE{Lense+Thirring1918,
       author = {{Lense}, Josef and {Thirring}, Hans},
        title = "{{\"U}ber den Einflu{\ss} der Eigenrotation der Zentralk{\"o}rper auf die Bewegung der Planeten und Monde nach der Einsteinschen Gravitationstheorie}",
      journal = {Physikalische Zeitschrift},
         year = 1918,
        month = jan,
       volume = {19},
        pages = {156},
       adsurl = {https://ui.adsabs.harvard.edu/abs/1918PhyZ...19..156L}
}

@ARTICLE{Antoniadis+2011,
       author = {{Antoniadis}, J. and {Bassa}, C.~G. and {Wex}, N. and {Kramer}, M. and {Napiwotzki}, R.},
        title = "{A white dwarf companion to the relativistic pulsar PSR J1141-6545}",
      journal = {\mnras},
         year = 2011,
        month = mar,
       volume = {412},
       number = {1},
        pages = {580-584},
          doi = {10.1111/j.1365-2966.2010.17929.x},
archivePrefix = {arXiv},
       eprint = {1011.0926},
 primaryClass = {astro-ph.SR},
       adsurl = {https://ui.adsabs.harvard.edu/abs/2011MNRAS.412..580A}
}

@ARTICLE{Venkatraman+2020,
       author = {{Venkatraman Krishnan}, V. and {Bailes}, M. and {van Straten}, W. and {Wex}, N. and {Freire}, P.~C.~C. and {Keane}, E.~F. and {Tauris}, T.~M. and {Rosado}, P.~A. and {Bhat}, N.~D.~R. and {Flynn}, C. and {Jameson}, A. and {Os{\l}owski}, S.},
        title = "{Lense-Thirring frame dragging induced by a fast-rotating white dwarf in a binary pulsar system}",
      journal = {Science},
         year = 2020,
        month = jan,
       volume = {367},
       number = {6477},
        pages = {577-580},
          doi = {10.1126/science.aax7007},
archivePrefix = {arXiv},
       eprint = {2001.11405},
 primaryClass = {astro-ph.HE},
       adsurl = {https://ui.adsabs.harvard.edu/abs/2020Sci...367..577V}
}

@ARTICLE{Srinivasan+vandenHeuvel1982,
       author = {{Srinivasan}, G. and {van den Heuvel}, E.~P.~J.},
        title = "{Some constraints on the evolutionary history of the binary pulsar PSR1913+16.}",
      journal = {\aap},
         year = 1982,
        month = apr,
       volume = {108},
        pages = {143-147},
       adsurl = {https://ui.adsabs.harvard.edu/abs/1982A&A...108..143S}
}

@ARTICLE{Church+2006,
       author = {{Church}, Ross P. and {Bush}, Stephanie J. and {Tout}, Christopher A. and {Davies}, Melvyn B.},
        title = "{Detailed models of the binary pulsars J1141-6545 and B2303+46}",
      journal = {\mnras},
         year = 2006,
        month = oct,
       volume = {372},
       number = {2},
        pages = {715-727},
          doi = {10.1111/j.1365-2966.2006.10897.x},
       adsurl = {https://ui.adsabs.harvard.edu/abs/2006MNRAS.372..715C}
}

@ARTICLE{Smarr+Blandford1976,
       author = {{Smarr}, L.~L. and {Blandford}, R.},
        title = "{The binary pulsar: physical processes, possible companions, and evolutionary histories.}",
      journal = {\apj},
         year = 1976,
        month = jul,
       volume = {207},
        pages = {574-588},
          doi = {10.1086/154524},
       adsurl = {https://ui.adsabs.harvard.edu/abs/1976ApJ...207..574S}
}

@ARTICLE{Lai+1995,
       author = {{Lai}, Dong and {Bildsten}, Lars and {Kaspi}, Victoria M.},
        title = "{Spin-Orbit Interactions in Neutron Star/Main-Sequence Binaries and Implications for Pulsar Timing}",
      journal = {\apj},
         year = 1995,
        month = oct,
       volume = {452},
        pages = {819},
          doi = {10.1086/176350},
archivePrefix = {arXiv},
       eprint = {astro-ph/9505042},
 primaryClass = {astro-ph},
       adsurl = {https://ui.adsabs.harvard.edu/abs/1995ApJ...452..819L}
}

@ARTICLE{Liu+2012,
       author = {{Liu}, K. and {Keane}, E.~F. and {Lee}, K.~J. and {Kramer}, M. and {Cordes}, J.~M. and {Purver}, M.~B.},
        title = "{Profile-shape stability and phase-jitter analyses of millisecond pulsars}",
      journal = {\mnras},
         year = 2012,
        month = feb,
       volume = {420},
       number = {1},
        pages = {361-368},
          doi = {10.1111/j.1365-2966.2011.20041.x},
archivePrefix = {arXiv},
       eprint = {1110.4759},
 primaryClass = {astro-ph.HE},
       adsurl = {https://ui.adsabs.harvard.edu/abs/2012MNRAS.420..361L}
}

@ARTICLE{Wang+vanLeeuwn2025,
       author = {{Wang}, Yuyang and {van Leeuwen}, Joeri},
        title = "{A deep search for radio pulsations from the 1.3 M$_{{\ensuremath{\odot}}}$ compact-object binary companion of young pulsar PSR J1906+0746}",
      journal = {\aap},
         year = 2025,
        month = sep,
       volume = {701},
          eid = {A180},
        pages = {A180},
          doi = {10.1051/0004-6361/202555920},
archivePrefix = {arXiv},
       eprint = {2507.17641},
 primaryClass = {astro-ph.HE},
       adsurl = {https://ui.adsabs.harvard.edu/abs/2025A&A...701A.180W}
}

@ARTICLE{Edwards+2006,
       author = {{Edwards}, R.~T. and {Hobbs}, G.~B. and {Manchester}, R.~N.},
        title = "{TEMPO2, a new pulsar timing package - II. The timing model and precision estimates}",
      journal = {\mnras},
         year = 2006,
        month = nov,
       volume = {372},
       number = {4},
        pages = {1549-1574},
          doi = {10.1111/j.1365-2966.2006.10870.x},
archivePrefix = {arXiv},
       eprint = {astro-ph/0607664},
 primaryClass = {astro-ph},
       adsurl = {https://ui.adsabs.harvard.edu/abs/2006MNRAS.372.1549E}
}

@INPROCEEDINGS{DuPlain+2008,
       author = {{DuPlain}, Ron and {Ransom}, Scott and {Demorest}, Paul and {Brandt}, Patrick and {Ford}, John and {Shelton}, Amy L.},
        title = "{Launching GUPPI: the Green Bank Ultimate Pulsar Processing Instrument}",
    booktitle = {Advanced Software and Control for Astronomy II},
         year = 2008,
       editor = {{Bridger}, Alan and {Radziwill}, Nicole M.},
       series = {Society of Photo-Optical Instrumentation Engineers (SPIE) Conference Series},
       volume = {7019},
        month = aug,
          eid = {70191D},
        pages = {70191D},
          doi = {10.1117/12.790003},
       adsurl = {https://ui.adsabs.harvard.edu/abs/2008SPIE.7019E..1DD}
}

@ARTICLE{Schlegel+1998,
       author = {{Schlegel}, David J. and {Finkbeiner}, Douglas P. and {Davis}, Marc},
        title = "{Maps of Dust Infrared Emission for Use in Estimation of Reddening and Cosmic Microwave Background Radiation Foregrounds}",
      journal = {\apj},
         year = 1998,
        month = jun,
       volume = {500},
       number = {2},
        pages = {525-553},
          doi = {10.1086/305772},
archivePrefix = {arXiv},
       eprint = {astro-ph/9710327},
 primaryClass = {astro-ph},
       adsurl = {https://ui.adsabs.harvard.edu/abs/1998ApJ...500..525S}
}

@ARTICLE{Rigby+2019,
       author = {{Rigby}, A.~J. and {Moore}, T.~J.~T. and {Eden}, D.~J. and {Urquhart}, J.~S. and {Ragan}, S.~E. and {Peretto}, N. and {Plume}, R. and {Thompson}, M.~A. and {Currie}, M.~J. and {Park}, G.},
        title = "{CHIMPS: physical properties of molecular clumps across the inner Galaxy}",
      journal = {\aap},
         year = 2019,
        month = dec,
       volume = {632},
          eid = {A58},
        pages = {A58},
          doi = {10.1051/0004-6361/201935236},
archivePrefix = {arXiv},
       eprint = {1909.04714},
 primaryClass = {astro-ph.GA},
       adsurl = {https://ui.adsabs.harvard.edu/abs/2019A&A...632A..58R}
}

@ARTICLE{Salaris+2022,
       author = {{Salaris}, Maurizio and {Cassisi}, Santi and {Pietrinferni}, Adriano and {Hidalgo}, Sebastian},
        title = "{The updated BASTI stellar evolution models and isochrones - III. White dwarfs}",
      journal = {\mnras},
         year = 2022,
        month = feb,
       volume = {509},
       number = {4},
        pages = {5197-5208},
          doi = {10.1093/mnras/stab3359},
archivePrefix = {arXiv},
       eprint = {2111.09285},
 primaryClass = {astro-ph.SR},
       adsurl = {https://ui.adsabs.harvard.edu/abs/2022MNRAS.509.5197S}
}

@ARTICLE{Schlafly+Finkelbeiner2011,
       author = {{Schlafly}, Edward F. and {Finkbeiner}, Douglas P.},
        title = "{Measuring Reddening with Sloan Digital Sky Survey Stellar Spectra and Recalibrating SFD}",
      journal = {\apj},
         year = 2011,
        month = aug,
       volume = {737},
       number = {2},
          eid = {103},
        pages = {103},
          doi = {10.1088/0004-637X/737/2/103},
archivePrefix = {arXiv},
       eprint = {1012.4804},
 primaryClass = {astro-ph.GA},
       adsurl = {https://ui.adsabs.harvard.edu/abs/2011ApJ...737..103S}
}

@ARTICLE{Hulse+Taylor1975,
       author = {{Hulse}, R.~A. and {Taylor}, J.~H.},
        title = "{Discovery of a pulsar in a binary system.}",
      journal = {\apjl},
         year = 1975,
        month = jan,
       volume = {195},
        pages = {L51-L53},
          doi = {10.1086/181708},
       adsurl = {https://ui.adsabs.harvard.edu/abs/1975ApJ...195L..51H}
}

@ARTICLE{Taylor+Weisberg1982,
       author = {{Taylor}, J.~H. and {Weisberg}, J.~M.},
        title = "{A new test of general relativity - Gravitational radiation and the binary pulsar PSR 1913+16}",
      journal = {\apj},
         year = 1982,
        month = feb,
       volume = {253},
        pages = {908-920},
          doi = {10.1086/159690},
       adsurl = {https://ui.adsabs.harvard.edu/abs/1982ApJ...253..908T}
}

@ARTICLE{Freire+Wex2024,
       author = {{Freire}, Paulo C.~C. and {Wex}, Norbert},
        title = "{Gravity experiments with radio pulsars}",
      journal = {Living Reviews in Relativity},
         year = 2024,
        month = dec,
       volume = {27},
       number = {1},
          eid = {5},
        pages = {5},
          doi = {10.1007/s41114-024-00051-y},
archivePrefix = {arXiv},
       eprint = {2407.16540},
 primaryClass = {gr-qc},
       adsurl = {https://ui.adsabs.harvard.edu/abs/2024LRR....27....5F}
}

@ARTICLE{Eardley1975,
       author = {{Eardley}, D.~M.},
        title = "{Observable effects of a scalar gravitational field in a binary pulsar.}",
      journal = {\apjl},
         year = 1975,
        month = mar,
       volume = {196},
        pages = {L59-L62},
          doi = {10.1086/181744},
       adsurl = {https://ui.adsabs.harvard.edu/abs/1975ApJ...196L..59E}
}

@ARTICLE{Damour+Esposito1996,
       author = {{Damour}, Thibault and {Esposito-Far{\`e}se}, Gilles},
        title = "{Tensor-scalar gravity and binary-pulsar experiments}",
      journal = {\prd},
         year = 1996,
        month = jul,
       volume = {54},
       number = {2},
        pages = {1474-1491},
          doi = {10.1103/PhysRevD.54.1474},
archivePrefix = {arXiv},
       eprint = {gr-qc/9602056},
 primaryClass = {gr-qc},
       adsurl = {https://ui.adsabs.harvard.edu/abs/1996PhRvD..54.1474D}
}

@ARTICLE{Bhat+2008,
       author = {{Bhat}, N.~D. Ramesh and {Bailes}, Matthew and {Verbiest}, Joris P.~W.},
        title = "{Gravitational-radiation losses from the pulsar white-dwarf binary PSR J1141 6545}",
      journal = {\prd},
         year = 2008,
        month = jun,
       volume = {77},
       number = {12},
          eid = {124017},
        pages = {124017},
          doi = {10.1103/PhysRevD.77.124017},
archivePrefix = {arXiv},
       eprint = {0804.0956},
 primaryClass = {astro-ph},
       adsurl = {https://ui.adsabs.harvard.edu/abs/2008PhRvD..77l4017B}
}

@ARTICLE{Freire+2012,
       author = {{Freire}, Paulo C.~C. and {Wex}, Norbert and {Esposito-Far{\`e}se}, Gilles and {Verbiest}, Joris P.~W. and {Bailes}, Matthew and {Jacoby}, Bryan A. and {Kramer}, Michael and {Stairs}, Ingrid H. and {Antoniadis}, John and {Janssen}, Gemma H.},
        title = "{The relativistic pulsar-white dwarf binary PSR J1738+0333 - II. The most stringent test of scalar-tensor gravity}",
      journal = {\mnras},
         year = 2012,
        month = jul,
       volume = {423},
       number = {4},
        pages = {3328-3343},
          doi = {10.1111/j.1365-2966.2012.21253.x},
archivePrefix = {arXiv},
       eprint = {1205.1450},
 primaryClass = {astro-ph.GA},
       adsurl = {https://ui.adsabs.harvard.edu/abs/2012MNRAS.423.3328F}
}

%\begin{thebibliography}{99}
%\end{thebibliography}

%%%%%%%%%%%%%%%%%%%%%%%%%%%%%%%%%%%%%%%%%%%%%%%%%%

%%%%%%%%%%%%%%%%% APPENDICES %%%%%%%%%%%%%%%%%%%%%

\begin{comment}

\appendix

\section{Some extra material}

If you want to present additional material which would interrupt the flow of the main paper,
it can be placed in an Appendix which appears after the list of references.
\end{comment}
%%%%%%%%%%%%%%%%%%%%%%%%%%%%%%%%%%%%%%%%%%%%%%%%%%

% Don't change these lines
\bsp	% typesetting comment
\label{lastpage}
\end{document}